%% file: main.tex
\documentclass{article}

\usepackage{graphicx}
\usepackage[hang]{subfigure}
\usepackage{url}
\usepackage{amsmath}
\usepackage{amsfonts}
\usepackage{amssymb}
\usepackage{amsthm}
\usepackage{color}
\usepackage{paralist}
\usepackage{ifthen}
\usepackage{array}
\usepackage{cite}
\usepackage{epstopdf}
\usepackage{epsfig}
\usepackage{enumerate}
\usepackage{footmisc}
\usepackage{amsbsy}
\usepackage{bm}
\usepackage{array,cases}
\usepackage{multirow}
\usepackage{algorithm}
\usepackage{algorithmic}
\usepackage{comment}
\usepackage{caption}
\usepackage{times}
\usepackage{authblk}

\newboolean{longver}
\setboolean{longver}{true}

\input{mymath}

\newtheorem{lemma}{Lemma}[section]

\newtheorem{theorem}{Theorem}

\newtheorem{definition}{Definition}

\newcommand\defeq{\mathrel{:=}}

\newcommand\ie{{\em i.e.}}
\newcommand\eg{{\em e.g.}}

\title{Simulation-based Distributed Coordination Maximization over Networks}
\author{Hyeryung~Jang, Jinwoo~Shin, and Yung~Yi
  \thanks{
Part of this paper was presented at ACM MOBIHOC, 2016\cite{jang:coordmax}. 
} 
\thanks{H. Jang is with the Department of Informatics, King's College London, London WC2R2LS, United Kingdom (e-mail: hyeryung.jang@kcl.ac.uk). J. Shin and Y. Yi are with the Department of Electrical Engineering, Korea Advanced Institute of Science and Technology, Daejeon 34141, South Korea (e-mails: jinwoos@kaist.ac.kr; yiyung@kaist.edu).} 
}

\date{}

\allowdisplaybreaks

\begin{document}
\maketitle

\newcommand{\expectation}{\textsf{E}}
\newcommand{\probability}{\textsf{P}}
\newcommand{\pdf}{\textsf{f}}
\newcommand{\slow}{\ell}
\newcommand{\nextline}{\mbox{}\\}
\newcommand{\ud}{\mathrm{d}}
\newcounter{tempcounter}
\newcounter{acounter}

\begin{abstract}
  In various online/offline multi-agent networked environments, it is very popular that the system can benefit from coordinating actions of two interacting agents at some cost of coordination. In this paper, we first formulate an optimization problem that captures the amount of coordination gain at the cost of node activation over networks. This problem is challenging to solve in a distributed manner, since the target gain is a function of the long-term time portion of the inter-coupled activations of two adjacent nodes, and thus a standard Lagrange duality theory is hard to apply to obtain a distributed decomposition as in the standard Network Utility Maximization. In this paper, we propose three simulation-based distributed algorithms, each having different update rules, all of which require only one-hop message passing and locally-observed information. The key idea for being distributedness is due to a stochastic approximation method that runs a Markov chain simulation incompletely over time, but provably guarantees its convergence to the optimal solution. Next, we provide a game-theoretic framework to interpret our proposed algorithms from a different perspective. We artificially select the payoff function, where the game's Nash equilibrium is asymptotically equal to the socially optimal point, \ie, no Price-of-Anarchy. We show that two stochastically-approximated variants of standard game-learning dynamics overlap with two algorithms developed from the optimization perspective. Finally, we demonstrate our theoretical findings on convergence, optimality, and further features such as a trade-off between efficiency and convergence speed through extensive simulations.
\end{abstract}

\input{intro}
\input{related}
\input{model}

\input{algorithm}
\input{approaches}  
\input{simulations}
\input{conclusion}

\appendix
\input{proofs}
\input{appendix}

\bibliographystyle{IEEEtran}
\bibliography{ref}

\end{document}

%% file: mymath.tex
\newcommand{\set}[1]{\ensuremath{\mathcal #1}}

\newcommand{\separator}{
  \begin{center}
    \rule{\columnwidth}{0.3mm}
  \end{center}
}

\newcommand{\beq}{\begin{eqnarray*}}
\newcommand{\eeq}{\end{eqnarray*}}
\newcommand{\beqn}{\begin{eqnarray}}
\newcommand{\eeqn}{\end{eqnarray}}
\newcommand{\bemn}{\begin{multiline}}
\newcommand{\eemn}{\end{multiline}}

\newcommand{\grad}[1]{\nabla #1}



%% file: intro.tex
\section{Introduction}
\label{sec:intro}

In many online/offline networking environments, a variety of gains
among nodes (or agents) are generated when they make efforts to adjust
their states (or actions) with those of others. Two examples include
the ones in wireless and social networks. First, in wireless sensor
networks with duty cycled node activations for energy saving, each
sensor node decides to be awake or not over time, which further
depends on its neighbors' wake-up state and distance to the node. When
two nearby nodes communicate, they are equipped with a robust wireless
channel for mutual communication, and thus their coordination (\eg,
message exchange) can become more powerful at the cost of energy
consumption while they are awake \cite{bulusu:coord, chen:span}. Thus,
to achieve the desired coordination gain while turning off redundant
sensors, each sensor node should smartly decide to wake up or not,
which should often be done in a distributed manner. Second, in
online/offline social networks, social relationships and interactions
are of critical interest, since the strength of such interactions
often determines how the network evolves, \eg, adoption of a new
technology or spread of information. For example, when a new
technology becomes available, using the social relationships, more
coordination gain due to the compatibility of the technology between
two individuals is generated, whereas a certain cost due to the
adoption of the new technology is incurred, \eg, buying a new software
\cite{montanari:spread, laciana:ising, bakshy:social}.

In this paper, we formulate an optimization problem, called {\em
  Coordination Gain Maximization} that suitably captures the gain due
to peer-to-peer coordinations of connected node pairs and the cost due
to individual node activations, as in the following form:
\begin{eqnarray}
\label{eq:coord_opt}
\max_{\lambda_i, \lambda_{ij}} 
  \sum_{\text{connected node pair } (i,j)} U_{ij}(\lambda_{ij}) - \sum_{\text{node } i} C_i(\lambda_i),
\end{eqnarray}
where $U_{ij}(\cdot)$ and $C_i(\cdot)$ are the coordination utility
and the node activation cost functions, respectively. Intuitively,
$\lambda_{ij}$ is the long-term time portion when {\em both} nodes $i$
and $j$ are simultaneously activated and thus coordinated, and
$\lambda_i$ is the long-term time portion when node $i$ is
activated. This optimization seems a simple variant of a standard NUM
(Network Utility Maximization) \cite{palomar:num, kelly:num, li:num},
where it is allowed to easily develop a node-wise distributed
algorithm converging to the optimal solution. However, the problem in
\eqref{eq:coord_opt} significantly differs from a standard NUM
problem, thus developing a distributed algorithm is far from being
trivial. The main challenge of solving this optimization problem lies
in the fact that the standard Lagrange duality theory for a
distributed decomposition is not possible since the objective function
includes the term which is a function of the long-term inter-coupling
of the states of a pair of connected nodes, and thus, a separability
is not permitted.

In many engineering systems, we often observe the trade-off between
efficiency and complexity, where optimal algorithms require extensive
message passing or heavy computations, but light-weight approximate
algorithms incur suboptimality. Stochastic {\em simulation-based} algorithms \cite{andrad98simul, gosavi03simul} have been investigated in various areas to handle expensive computations in efficient way by using random experimental simulations, in spite of some challenges such as slow convergence time and/or suboptimality of the resulting solution. Our primary goal is to develop a simulation-based distributed coordination decision algorithm that is ``efficient'',
\ie, hopefully achieving the optimal solution of \eqref{eq:coord_opt} using random samples of configurations produced by locally-limited message passing. In this work, we formulate an
optimization problem of coordination gain maximization over networks
by taking a framework of the binary pairwise undirected graphical
model, \ie, Ising model \cite{ising:ising, mccoy:ising}, to capture
pairwise coordinations and nodewise activations of the network, and then propose distributed dynamic mechanisms which produce the optimal solution of \eqref{eq:coord_opt}. Our
main contributions are summarized in what follows:

\smallskip
\noindent{\bf \em C1.} We first introduce a distributed mechanism,
called {\bf CDM($\bm{\theta}$)} (Configuration Decision Mechanism),
where, by each node's local state changes based on one-hop message
passing, a node activation state of the network is randomly determined
in a decentralized manner. This mechanism is governed by a given
parameter vector $\bm{\theta}$ that represents the strength of
inter-node coordinations and the preference for activation of each
node.  We illustrate how {\bf CDM} aids in the design of distributed,
efficient coordination decision algorithm.

\smallskip
\noindent {\bf \em C2.} We then propose three simulation-based
algorithms, called {\bf \em Coord-dual}, {\bf \em Coord-steep}, and
{\bf \em Coord-ind} that provide how to update the parameter vector
$\bm{\theta}$ of {\bf CDM} in a distributed manner. We prove that all
of three algorithms provably converge to the optimal solution of
\eqref{eq:coord_opt}, yet the rationale behind each scheme contains
different perspectives of approximation and optimization
mechanisms. The key technique towards a distributed operation is to
run {\bf CDM($\bm{\theta}$)} incompletely over time and exploit {\em
  locally-observed} information from random samples to update the
parameter $\bm \theta$, which can guarantee the convergence to the
optimal solution of \eqref{eq:coord_opt} on the strength of stochastic
approximation theory.

\smallskip
\noindent {\bf \em C3.} Finally, we take a different angle to
understand two algorithms {\bf \em Coord-steep} and {\bf \em
  Coord-ind} using game theory. A game-theoretic framework is one of
the powerful tools in the design and analyze the behavior of
multi-agent systems, providing valuable insights into various local
control rules for agents' behaviors \cite{NJ13}. In this paper, we
design a non-cooperative game with {\em artificially-selected}
payoffs, and show that it has a unique Nash equilibrium which is
(asymptotically) equivalent to the socially optimal point, \ie, the
optimal solution of \eqref{eq:coord_opt}. We consider popular game
dynamics, which we modify with the stochastic approximation technique,
and find that those two game dynamics exactly correspond to {\bf \em
  Coord-steep} and {\bf \em Coord-ind}, respectively. We conduct
extensive simulations to verify our theoretical findings.

\smallskip
\noindent{\bf \em Organization.} The rest of the paper is organized as
follows. In Section~\ref{sec:related}, we present a large array of
related works. In Section~\ref{sec:model}, we formulate the coordination gain maximization problem, followed
by the analysis of distributed coordination algorithms in
Section~\ref{sec:alg_analysis}. In Section~\ref{sec:interpretation}, we provide interpretations from a
game-theoretic framework, demonstrate the performance of our
algorithms through numerical results in Section~\ref{sec:sim}, and
finally conclude in Section~\ref{sec:conclusion}. All the mathematical
proofs are presented in Appendix.


%% file: related.tex
\section{Related Work}
\label{sec:related}

A variety of benefits from coordinating actions of wireless terminals
or users have been widely studied in wireless networks. In the area of
wireless sensor networks, various distributed, energy-efficient
coordination schemes have been proposed recently, where sensors
adaptively select to be coordinators or not, \ie, stay awake and
forward sensing data or not, while turning off redundant sensors for
energy efficiency. The main interest of this area is scalable,
localized, and robust coordination in large-scale environments, and
thus most works have been studied to (i) preserve capacity and
connectivity \cite{chen:span}, (ii) improve the lifetime of the system
and communication latency by using a geo-location information of
sensors \cite{xu:gat}, or (iii) build a self-configuring localization
system \cite{bulusu:coord}.

In online/offline social networks, coordinating actions, \eg,
diffusion /adoption of information, of two (socially) interacting
individuals is of importance, since the power of interactions often
determines how the network evolves. An importance of a coordination
mechanism, \eg, a social structure, for efficient knowledge sharing
has been stressed in \cite{tsai:social}. There has been a surge of
studies about the dynamics of diffusion (\ie, a status of agents) by
adopting Ising model in statistical physics \cite{laciana:ising},
epidemic-based models \cite{kempe:spread}, or game-theoretic models
\cite{montanari:spread}. Recently, researchers have studied how
coordinated behavior might spread in a network, \ie, game-theoretic
diffusion models, and the impact of network structure and/or seeding
set on convergence speed \cite{kandori:learning, ellison:coordination,
  montanari:convergence, liu:influence}. Especially, the authors in
\cite{kandori:learning} studied that the noisy best response dynamics
converges to the equilibrium, which maximizes the spreading efficiency
by choosing an appropriate seed set as in \cite{liu:influence}.

A large array of work about network utility maximization (NUM) problem
have been studied, see \cite{palomar:num, kelly:num, li:num} for
surveys. The objective of NUM problem is to maximize a sum of all
nodes' utilities, while not considering any pairwise status, thus
separability in the problem often provides a useful dual-based
decomposition for an easy development of distributed algorithms. In
recent years, the researches on achieving optimality in both
throughput and utility in wireless scheduling (in a decentralized
manner) have been studied from an optimization perspective
\cite{jiang:ocsma, liu:uocsma} as well as from game-theoretic
perspective \cite{chen2007game, cui:game_mac, jang:ocsma} for various
base-line medium access control protocols. The intuitive idea of these
works is that wireless links adaptively adjust access intensities by
using local information, \eg, queue-length or empirical service rate,
so as to achieve the desired performance.

Our work is based on the importance of the pairwise coordination
impacts among individuals, where our main interest is how to find a
sequence of node activations (and thus coordinations) in a {\em
  decentralized manner} whose long-term status leads to the solution
of the problem in \eqref{eq:coord_opt} that maximizes the network-wide
coordination gain at the cost of node activation. Moreover, our work
provides new interpretations behind the results obtained from a
game-theoretic perspective in the sense that (i) we start from a
non-cooperative (ordinal potential) game, followed by the resulting
Nash equilibriums' efficiency (\ie, no Price-of-Anarchy), and (ii) we
provide how game-inspired learning dynamics of the game can be
connected to the results from an optimization approach.


%% file: model.tex
\section{Model and Objective}
\label{sec:model}

\subsection{System Model}

\noindent {\bf \em Network model.} In this paper, we consider a
network $G=(V,E)$ consisting of a set $V$ of nodes and a set
$E \subset V \times V$ of edges. With this graphical representation,
each node corresponds to an agent in social networks or a sensor node
in wireless sensor networks, and each edge corresponds to a physical
connectivity or a social relationship between nodes, \ie,
$(i,j) \in E$ means that node $i$ and node $j$ are connected and have
an interaction. Note that we study undirected networks where
interaction requires mutual consent, \ie, $(i,j)$ is equivalent to
$(j,i)$. Let $\set{N}(i) = \{ j \in V: (i,j) \in E\}$ denote the
neighbors of node $i$.

\smallskip
\noindent {\bf \em Configuration and coordination scheduling.} We
consider a continuous time framework. Let $\sigma_i(\tau) \in \{0,1\}$
indicate whether node $i$ is {\em active} at time $\tau$ or not, \ie,
$\sigma_i(\tau)=1$ means that node $i$ is active at time $\tau$, and
$0$ otherwise. We say that nodes $i$ and $j$ are (or edge $(i,j)$ is)
{\em coordinated} when $\sigma_i(\tau)\sigma_j(\tau)=1$. We also
denote by ${\bm \sigma}(\tau) = [\sigma_i(\tau)]_{i \in V}$ a {\em
  node configuration} at time $\tau$, and it is clear that a set of
possible configurations of the graph $G$ is defined as
$\set{I}(G) \defeq \{0,1\}^{|V|}$. To formally discuss a coordination
gain, which we will introduce later, we define a {\em coordination
  configuration} as follows:
\begin{eqnarray} \label{eq:coorconf}
{\bm \phi}(\bm \sigma) \defeq ([\sigma_i]_{i \in V}, [\sigma_i\sigma_j]_{(i,j) \in E}),
\end{eqnarray}
which is an augmented configuration vector capturing {\em both} the
activation status of nodes and the coordination status of edges. Then,
every coordination configuration belongs to
$\Phi(G) \defeq \{0,1\}^{|V|+|E|}$. Now, a {\em coordination
  scheduling} (or simply scheduling) algorithm is a mechanism that
selects ${\bm \sigma}(\tau) \in \set{I}(G)$ (thus a coordination
configuration ${\bm \phi}({\bm \sigma}(\tau)) \in \Phi(G)$ is also
determined) over time $\tau \in \mathbb{R}_+$.

\smallskip
\noindent{\bf \em Coordination region.} We now define the maximum
achievable {\em coordination region} (also called coordination
capacity region) $\Lambda \subset [0,1]^{|V|+|E|}$ of the network,
which is the convex hull of the feasible coordination configuration
set $\Phi(G)$, \ie,
\begin{eqnarray*}
  \Lambda = \Lambda(G) \defeq \bigg\{ \sum_{\bm \sigma \in \set{I}(G)} {\mu}_{\bm \sigma}{\bm \phi}(\bm \sigma) 
  : \sum_{\bm{\sigma} \in \set{I}(G)} {\mu}_{\bm{\sigma}} = 1, {\mu}_{(\cdot)} \geq 0 \bigg\}.
\end{eqnarray*}
The intuition of the notion of coordination capacity region comes from
the fact that any coordination scheduling algorithm has to choose a
node configuration from $\set{I}(G)$ over time (thus a coordination
${\bm \phi}(\bm \sigma)$ is determined), and $w_{\bm \sigma}$ denotes
the fraction of time selecting a node configuration ${\bm \sigma}$
(and thus a coordination ${\bm \phi}(\bm \sigma)$). Hence, the
long-term (average) time portion of node activation and edge
coordination induced by any scheduling algorithm must belong to
$\Lambda$.

\subsection{Problem Description: Coordination Gain Maximization}
\noindent {\bf \em Objective.} We require nodes and edges to control
the long-term time portion (or frequency) of activation and
coordination close to some boundary of $\Lambda$. Specifically, we aim
at designing a coordination scheduling algorithm that decides
${\bm \sigma}(\tau) \in \set{I}(G)$ over time $\tau$ so that the
long-term time portion of node activation and edge coordination converges to a solution of the following optimization problem:
\vspace{-0.3cm} \separator \vspace{-0.25cm}
\begin{eqnarray} \label{eq:opt} 
\textbf{(CG-OPT)} \quad \max_{{\bm
      \lambda} \in \Lambda} \quad \sum_{(i,j) \in E}
  U_{ij}(\lambda_{ij}) - \sum_{i \in V} C_i(\lambda_i).
\end{eqnarray}
\vspace{-0.65cm}
\separator
\vspace{-0.15cm}

This problem, which we call {\em coordination gain maximization}
problem, captures inter-dependencies among nodes and a trade-off
between (edge) coordination utility and (node) activation cost, where
$U_{ij}: [0,1] \rightarrow \mathbb{R}$ is a strictly concave and continuously twice-differentiable coordination utility function of edge
$(i,j) \in E$, and $C_i: [0,1] \rightarrow \mathbb{R}$ is a strictly
convex, continuously twice-differentiable activation cost function of node
$i \in V$. Then, it is obvious that \textbf{CG-OPT} has the unique solution
${\bm \lambda}^\star \defeq ([\lambda^\star_i]_{i \in V},
[\lambda^\star_{ij}]_{(i,j) \in E})$. The network coordination gain is
defined as a total coordination utility subtracted by a total
incurring cost. More coordination utility is generated as nodes $i$
and $j$ are coordinated more often, but it also incurs more cost of
nodes $i$ and $j$ to be activated.

\begin{figure}[t!]
  \centering
  \includegraphics[width=0.35\columnwidth]{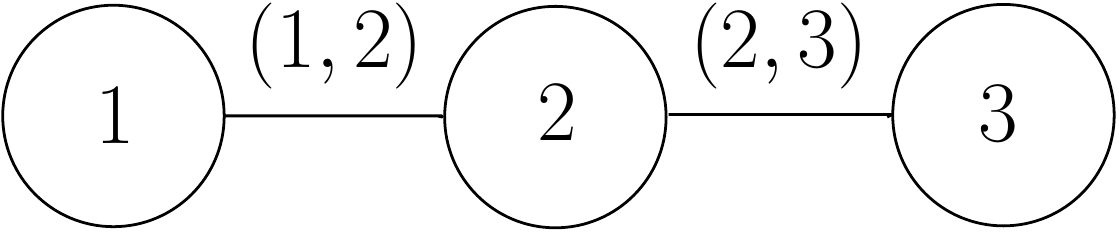}
  \caption{{An example line network with $3$ nodes and $2$ edges, where
    there are $8$ feasible node configurations ${\bm \sigma} \in
    \{0,1\}^3$.}}
  \label{fig:ex}
\end{figure}

\smallskip
\noindent{\bf \em Example.} To illustrate, we provide an example of
{\bf CG-OPT} and its solution structure, where we use a line topology
with $3$ nodes and $2$ edges, as depicted in Fig.~\ref{fig:ex}. Now,
{\bf CG-OPT} in this example is expressed by:
\begin{align*}
  \max_{{\bm \lambda}} \ \Bigg [U_{12}(\lambda_{12}) +
  U_{23}(\lambda_{23}) - \Big ( C_1(\lambda_1) + C_2(\lambda_2) +
  C_3(\lambda_3) \Big ) \Bigg].
\end{align*}
Let the long-term time portion of activation and coordination be
characterized by the distribution
$\{ \pi_{\bm \sigma} \}_{{\bm \sigma} \in \{0,1\}^3}$, \ie,
\begin{eqnarray} \label{eq:long-term}
\lambda_1 &=& \pi_{(1,0,0)}+\pi_{(1,0,1)}+\pi_{(1,1,0)}+\pi_{(1,1,1)}, \cr
\lambda_2 &=& \pi_{(0,1,0)}+\pi_{(0,1,1)}+\pi_{(1,1,0)}+\pi_{(1,1,1)}, \cr
\lambda_3 &=& \pi_{(0,0,1)}+\pi_{(0,1,1)}+\pi_{(1,0,1)}+\pi_{(1,1,1)}, \cr
\lambda_{12} &=& \pi_{(1,1,0)}+\pi_{(1,1,1)}, ~\mbox{}~
\lambda_{23} = \pi_{(0,1,1)}+\pi_{(1,1,1)}.
\end{eqnarray}

Note that the total coordination gain is generated according to the
long-term coordination time portion of two edges, \ie,
$\lambda_{12}, \lambda_{23}$, and the total incurring cost is due to
the long-time activation of three nodes, \ie,
$\lambda_1, \lambda_2, \lambda_3$.

A smart scheduling is required since each node's activation should be
coordinated with its neighboring nodes in order to produce enough gain
at the cost of activation. For expositional convenience, let us choose
the following utility and cost functions:
$U_{12}(x)=U_{23}(x) = \log(x)$, $C_1(x) = C_2(x) = x^2$, and
$C_3(x) = 3x^2$, \ie, more cost is incurred for node $3$. Now a simple
algebra gives the following distributions and the resulting optimal
solution:
$$\pi^\star_{(0,0,0)} = 0.5, \ \pi^\star_{(1,1,0)} = 0.0915, \ \pi^\star_{(1,1,1)} =
0.4085,$$
$$(\lambda_1^\star,\lambda_2^\star,\lambda_3^\star,\lambda_{12}^\star,\lambda_{23}^\star)
= (0.5,0.5,0.4085,0.5,0.4085),$$ where the optimal solution is
attained by assigning some probability to the configuration $(1,1,0)$
rather than giving a high priority only to $(1,1,1)$, with some cost
balancing by avoiding the activation of any node, \ie, scheduling
$(0,0,0)$.

\smallskip
In this work, our goal is to design a {\em distributed} coordination
scheduling algorithm $\{{\bm \sigma}(\tau)\}_{\tau = 0}^{\infty}$
which relies only on local information with one-hop message passing,
but converges to the optimal solution of \textbf{CG-OPT}, \ie, $\lim_{T\to \infty}\frac{1}{T}\int_{0}^{T} {\bm \phi}({\bm \sigma}(\tau)) \mathrm{d}\tau = {\bm \lambda}^\star.$ A lot of challenges may arise, because the developed algorithm should work in an independent manner of the underlying topology and the shape of utility/cost functions, and more importantly the solution should be found in a distributed way, which may entail heavy computations to characterize the long-term time portion of activation and coordination. To overcome these challenges in efficient way, we propose distributed simulation-based algorithms in next section.


%% file: algorithm.tex
\section{Distributed Coordination: Algorithm and Analysis} \label{sec:alg_analysis}

\subsection{Configuration Decision Mechanism} \label{sec:cdm}

We first introduce a parameter ${\bm \theta} \in \mathbb{R}^{|V|+|E|}$
as:
\begin{eqnarray} \label{eq:param}
{\bm \theta} = ([\theta_i]_{i \in V},[\theta_{ij}]_{(i,j) \in E}).
\end{eqnarray}
Intuitively, $\theta_{ij}$ and $\theta_i$ can be interpreted as the
{\em strength} of coordination interaction of edge $(i,j) \in E$ and
the {\em preference} for activation of node $i \in V$,
respectively. To capture pairwise interaction of the system, the
parameter includes singleton as well as pairwise element, and this
parameter will be used as a parameter of algorithms we will design in
Section~\ref{sec:alg}.

Note that a coordination gain and activation cost of the system would
be a function of the long-term time portion of edge coordinations and
node activations (\ie, a stationary distribution, say $\pi$, of
configurations, see the example in \eqref{eq:long-term}), as hinted in
{\bf CG-OPT}. This means that it is necessary to develop a
time-by-time dynamic mechanism, which, if run for a sufficient amount
of time, leads to a certain stationary distribution of configurations
for a given $\bm \theta$. In this section, we illustrate how a simple
Monte Carlo Markov Chain method may be used as such a time-by-time
dynamic mechanism, called {\bf CDM} (Configuration Decision
Mechanism). We then identify the optimal distribution over the
feasible configurations that maximizes the network-wide coordination
gain by producing random samples of configurations via {\bf CDM}. Each
algorithm we propose in Section~\ref{sec:alg} produces a sequence of
configurations $\{ {\bm \sigma}(\tau)\}_{\tau=0}^\infty$ by updating a
parameter $\bm \theta$ over time so that the resulting long-term
activation/coordination rate converges to the optimal solution of {\bf
  CG-OPT}.

\begin{figure}[ht!]
\caption*{}
\vspace{-0.8cm}
\separator
\vspace{-0.2cm}
Configuration Decision Mechanism: \textbf{CDM}($\bm \theta$)
\vspace{-0.4cm}
\separator
\vspace{-0.2cm}
\textbf{Input:} Parameter $\bm \theta$, current configuration ${\bm \sigma}=[\sigma_i]$.\\
\textbf{Output:} New configuration ${\bm \sigma}'=[\sigma'_i]$.\\
\vspace{-0.1cm} \hrule 
\vspace{0.3cm}
Each node, say $i,$ when its Poisson clock ticks, performs the following:
\begin{compactenum}[\bf S1.]
\item Node $i$ changes its configuration from $\sigma_i$ to
  $\sigma_i'$
\begin{eqnarray} \label{eq:gd}
\sigma'_i = 
\begin{cases} 
  1, & \mbox{with probability } \frac{\exp(\theta_i + \sum_{j \in \set{N}(i)}\sigma_j \theta_{ij})}{1 + \exp(\theta_i + \sum_{j \in \set{N}(i)}\theta_{ij}\sigma_j)} \\
  0, & \mbox{with probability } \frac{1}{1 + \exp(\theta_i +
    \sum_{j \in \set{N}(i)}\theta_{ij}\sigma_j)}
\end{cases}
\end{eqnarray}

\item Node $i$ broadcasts its updated $\sigma'_i$ to all of neighbors in $\set{N}(i).$
\end{compactenum}
\vspace{-0.4cm}
\separator
\end{figure}

We now describe $\textbf{CDM}(\bm \theta)$ for a given parameter
$\bm \theta$, where every node has a Poisson clock with unit rate and
nodes decide a new configuration ${\bm \sigma}'$ from a current
configuration $\bm \sigma$ by the procedures as in two steps {\bf S1}
and {\bf S2}. Note that Poisson clock of each node leads to the
uniform node selection, and for a given graph $G$, {\bf CDM} decides a
configuration over time in a distributed manner with only {\em one-hop
  message passing}. In particular, when node $i$'s clock ticks, it requires to know
  (i) configuration status of its neighboring nodes, \ie, $\{
  \sigma_j\}_{j \in \set{N}(i)}$, and (ii) parameter of its
  neighboring edges, \ie, $\{\theta_{ij}\}_{j \in \set{N}(i)}$, to
  decide its new configuration $\sigma_i'$ in \textbf{S1}. Then, in
  \textbf{S2}, node $i$ broadcasts its updated configuration $\sigma_i'$ to all of
  its neighboring nodes $j \in \set{N}(i)$ for further configuration
  decision process. Simply, {\bf CDM($\bm \theta$)} decides a new
configuration mainly based on the status of neighbors with large
interaction strength.
  
One important feature is that \textbf{CDM} for a given parameter
$\bm \theta$ leads to a continuous-time Markov chain
$\{ {\bm \sigma}(\tau) \}_{\tau=0}^{\infty}$ achieving the following
stationary distribution
$p_{\bm \theta} = [p_{\bm \theta, \bm \sigma}]_{{\bm \sigma} \in
  \set{I}(G)}$: 
\begin{align} \label{eq:ccd-stationary}
  p_{\bm \theta, \bm \sigma} \propto \exp( \langle {\bm \theta}, {\bm \phi}(\bm \sigma) \rangle ), \quad \text{for} ~ {\bm \sigma} \in \set{I}(G), 
\end{align}
where $\langle {\bm a},{\bm b} \rangle$ is the inner product of two
vectors $\bm a$ and $\bm b$, \ie,
$p_{\bm \theta, \bm \sigma} \propto \exp (\sum_{i \in
  V}\theta_i\sigma_i + \sum_{(i,j) \in E}
\theta_{ij}\sigma_i\sigma_j)$. Moreover,
$\{{\bm \sigma}(\tau)\}_{\tau=0}^\infty$ is an irreducible, aperiodic,
and reversible Markov process \cite{levin:markov}. Given the parameter
$\bm \theta$, the ergodicity and reversibility of the Markov process
imply that the marginal probability of nodes and edges under the
stationary distribution $p_{\bm \theta}$, denoted by
${\bm s}(\bm \theta) = ([s_i(\bm \theta)]_{i \in V}, [s_{ij}(\bm
\theta)]_{(i,j) \in E})$, becomes the long-term time portion of node
activation and edge coordination, simply called
activation/coordination rate, characterized as\footnote{We
  interchangeably use a notation of
  $\mathbb{E}_{p_{\bm \theta}}[\cdot]$ and
  $\mathbb{E}_{\bm \theta}[\cdot]$ for the expectation over the
  distribution $p_{\bm \theta}$.}: for $i \in V$ and $(i,j) \in E$,
\begin{align} \label{eq:marginal}
  s_i(\bm \theta) &= \mathbb{E}_{p_{\bm \theta}}[\sigma_i] = \sum_{{\bm \sigma} \in \set{I}(G)} p_{\bm \theta, \bm \sigma} \sigma_i, \cr
  s_{ij}(\bm \theta) &= \mathbb{E}_{p_{\bm \theta}}[\sigma_i\sigma_j] = \sum_{{\bm \sigma} \in \set{I}(G)} p_{\bm \theta, \bm \sigma} \sigma_i \sigma_j.
\end{align}
We remark that a graphical model representing the distribution in
\eqref{eq:ccd-stationary} corresponds to {\em Ising model} in
statistic physics, with {\em Ising parameter} $\bm \theta$
\cite{ising:ising}, and $\textbf{CDM}(\bm \theta)$ is a Glauber
dynamics over an Ising model under continuous-time setting.

\subsection{Simulation-based Parameter Update
  Algorithms} \label{sec:alg}

In {\bf CG-OPT}, our goal is to find a distribution $\bm \mu$ over
configurations such that the resulting rate
$\mathbb{E}_{\bm \mu}[{\bm \phi}(\bm \sigma)]$ becomes the optimal
solution of {\bf CG-OPT}. To that end, using
\textbf{CDM}($\bm \theta$), we develop three {\em distributed}
simulation-based algorithms that adaptively update the parameter
$\bm \theta$ over time, which we call {\bf \em Coord}-algorithms,
whose empirical activation/coordination rates from samples generated
by \textbf{CDM}($\bm \theta$) asymptotically converge to the optimal
solution of {\bf CG-OPT}.

\begin{algorithm}[ht!] 
  \caption*{\textbf{\em Coord}-algorithms: At the start of each frame
    $t \in \mathbb{Z}_{\geq 0}$} \label{alg:coord}
  \begin{algorithmic} \vspace{0.05in}
    \STATE \textbf{Input:} Efficiency parameter $\beta > 0$, smoothing parameter $\alpha \in (0,1]$, boundary values $\theta^{\text{min}}, \theta^{\text{max}}$. \\
    \STATE \textbf{Output:} ${\bm \theta}[t+1]$. \\
    \STATE \textbf{Initialize:} Set ${\bm \theta}[0]$ arbitrarily, and $a[0] = 0.$ \\
    \vspace{0.1in} \hrule \vspace{0.1in}

    \STATE {\bf S1.} Each node $i$ sends $\theta_{ij}[t]$ to its neighbor       $j$ for all  $j \in \set{N}(i)$.
    
    \vspace{0.05in}

    \STATE {\bf S2.} \textbf{CDM}(${\bm \theta}[t]$) is run by each node
    in a distributed manner, and each node $i$
    records the number of its activations $\hat{s}_i[t]$ and its
    coordinations $\{\hat{s}_{ij}[t]\}_{j \in \set{N}(i)}$ over frame
    $t$, and compute the cumulative rate
    $\bar{s}_i[t], \{\bar{s}_{ij}[t]\}_{j \in \set{N}(i)}$, as in \eqref{eq:agg-instant}.
    \vspace{0.05in}
    \STATE {\bf S3.} Each node $i$ updates $\theta_i[t+1]$ and
    $\{ \theta_{ij}[t+1]\}_{j \in \set{N}(i)}$ as:

	\vspace{0.05in}
    {\bf (a) {\em Coord-dual}:} 
    \begin{align} \label{eq:Q-dual}
      \theta_i[t+1] &= \bigg[
        \theta_i[t] + a[t] \bigg(
        C_i'^{-1}\bigg(\frac{-\theta_i[t]}{\beta} \bigg) -
        \hat{s}_i[t] \bigg)
        \bigg]_{\theta^{\text{min}}}^{\theta^{\text{max}}}, \cr
       \theta_{ij}[t+1] &= \bigg[
        \theta_{ij}[t] + a[t] \bigg(
        U_{ij}'^{-1}\bigg(\frac{\theta_{ij}[t]}{\beta} \bigg) -
        \hat{s}_{ij}[t] \bigg)
        \bigg]_{\theta^{\text{min}}}^{\theta^{\text{max}}}.
    \end{align}
    
    {\bf (b) {\em Coord-steep}:} 
    \begin{align} \label{eq:Q-steep}
      \theta_i[t+1] &= \bigg[ \theta_i[t] + \alpha \bigg( -\beta C_i' \Big(
        \bar{s}_i[t] \Big) - \theta_i[t] \bigg)
        \bigg]_{\theta^{\text{min}}}^{\theta^{\text{max}}}, \cr
      \theta_{ij}[t+1] &= \bigg[ \theta_{ij}[t] + \alpha \bigg( \beta U_{ij}' \Big( \bar{s}_{ij}[t] \Big) - \theta_{ij}[t] \bigg) \bigg]_{\theta^{\text{min}}}^{\theta^{\text{max}}}.
    \end{align}    

  {\bf (c) {\em Coord-ind}:}
      \begin{align} \label{eq:Q-ind}
 \theta_i[t+1] &= \bigg[ \theta_i[t] + \frac{\alpha}{\beta}\frac{\partial s_{i}({\bm \theta}[t])}{\partial \theta_{i}} \bigg( -\beta C_i' \Big(
        \bar{s}_i[t] \Big) - \theta_i[t] \bigg)
   \bigg]_{\theta^{\text{min}}}^{\theta^{\text{max}}}, \cr
   \theta_{ij}[t+1] &= \bigg[ \theta_{ij}[t] + \frac{\alpha}{\beta}\frac{\partial s_{ij}({\bm \theta}[t])}{\partial \theta_{ij}} \bigg( \beta U_{ij}'
        \Big( \bar{s}_{ij}[t] \Big) - \theta_{ij}[t] \bigg)
      \bigg]_{\theta^{\text{min}}}^{\theta^{\text{max}}}.
      \end{align}
\end{algorithmic}
\end{algorithm}
\vspace{0.2cm}

\smallskip
We now describe {\bf \em Coord}-algorithms, where $\beta>0$,
$\alpha \in (0,1]$, $\theta^{\text{min}}, \theta^{\text{max}}$ are the
given constants; $[\cdot]_x^y \defeq \max(y,\min(x,\cdot))$; and
$a:\mathbb{Z}_{\geq 0} \rightarrow \mathbb{R}_+$ is a positive
step-size function. In {\bf \em Coord}-algorithms, time is divided
into frames $t=0,1,\cdots$ of fixed durations $T$, and each node $i$
updates the parameter $\theta_i$ and
$\{\theta_{ij}\}_{j \in \set{N}(i)}$\footnote{In practice, each node
  $i$ may have additional independent controllers for its neighboring
  edges $(i,j)$ of $j \in \set{N}(i)$. Either node $i$ or $j$ may have
  the control authority of edge $(i,j)$ following some arbitrary
  rule.} following one of three schemes: (a) {\bf \em Coord-dual}, (b)
{\bf \em Coord-steep} and (c) {\bf \em Coord-ind}. 
In \textbf{S1}, at the beginning of each frame $t$, node $i$
  sends each of $\{\theta_{ij}[t]\}_{j \in \set{N}(i)}$ to each $j$ of
  its neighbors\footnote{We implicitly assume that
    information exchange in \textbf{S1} of {\bf \em Coord}-algorithms
    and \textbf{S2} in $\textbf{CDM}({\bm \theta})$ can be done by
    out-of-band signaling, \ie, a separate control channel.}. Then in \textbf{S2}, $\textbf{CDM}({\bm \theta}[t])$
runs in a distributed manner, leading to the local computation of {\em
  instant} and {\em cumulative} activation/coordination rates at and
until the frame $t$, denoted by $\hat{\bm s}[t]$ and $\bar{\bm s}[t]$,
respectively, \ie, 
\begin{align} \label{eq:agg-instant} 
\hat{\bm s}[t] = \frac{1}{T}
  \int_{tT}^{(t+1)T} {\bm \phi}({\bm \sigma}(\tau)) \mathrm{d}\tau, \quad
  \bar{\bm s}[t] = \frac{1}{t} \sum_{m = 0}^{t} \hat{\bm s}[m],
\end{align}
where both empirical rates are {\em locally-computed}. 
In {\bf S3},
each scheme utilizes either of the computed empirical rates:
$\hat{\bm s}[t]$ for {\bf \em Coord-dual} and $\bar{\bm s}[t]$ for
{\bf \em Coord-steep, Coord-ind}. Note that
$\grad {\bm s}({\bm \theta}[t])$ is also locally obtained (see Appendix for a detailed form of $\grad {\bm s}(\cdot)$),
thus all of {\bf \em Coord}-algorithms are run in a distributed
manner.

\subsection{Rationale behind {\bf \em
    Coord}-algorithms} \label{sec:rationale}

We now explain the rationale behind each scheme of {\bf \em
  Coord}-algorithms that contains different perspectives of
approximation and optimization mechanisms.

\smallskip
\noindent{\bf (a) {\em Coord-dual:}} Note that {\bf CG-OPT} in
\eqref{eq:opt} can be written as:
\begin{eqnarray} \label{eq:opt2}
\max_{\bm \mu \in \set{M}} ~\mbox{} \set{F}(\bm \mu) \defeq \sum_{(i,j) \in E} U_{ij}( \mathbb{E}_{\bm \mu}[\sigma_i \sigma_j] ) - \sum_{i \in V} C_i( \mathbb{E}_{\bm \mu}[\sigma_i] ),
\end{eqnarray}
where $\set{M}$ is a set of all probability measures over the feasible
configurations $\set{I}(G)$. From this, we consider the following
variant {\bf A-CG-OPT} (parameterized by $\beta>0$) of {\bf CG-OPT}:

\vspace{-0.35cm} \separator \vspace{-0.4cm}
\begin{eqnarray} \label{eq:aopt} \text{\bf (A-CG-OPT)} \cr
  \max &&
  \sum_{(i,j) \in E} U_{ij}(\lambda_{ij}) -
  \sum_{i \in V} C_i(\lambda_i) + \frac{1}{\beta} H(\bm \mu) \cr \text{over} && \bm{\mu}
  \in \set{M}, ~ {\bm \lambda} \in [0,1]^{|V|+|E|} \cr
  ~\text{subject to} && \lambda_i = \mathbb{E}_{\bm{\mu}}[\sigma_i],
  \quad \forall i \in V, \cr && \lambda_{ij} =
  \mathbb{E}_{\bm{\mu}}[\sigma_i\sigma_j], \quad \forall (i,j) \in E,
\end{eqnarray}
where
$H(\bm{\mu}) = - \sum_{{\bm \sigma} \in \set{I}(G)} \mu_{\bm \sigma}
\log \mu_{\bm \sigma}$ is the {\em entropy} of ${\bm \mu}$.
\vspace{-0.25cm} \separator \vspace{-0.1cm}

Note that, compared to {\bf CG-OPT}, {\bf A-CG-OPT} has additional
term $\frac{1}{\beta} H(\bm \mu)$ in its objective function. Since the
entropy is bounded, \ie, $|H(\bm \mu)| \leq \log |\set{I}(G)|$, a solution of {\bf A-CG-OPT}, say
$({\bm \mu}^\circ, {\bm \lambda}^\circ)$, becomes arbitrarily closer
to that of {\bf CG-OPT} for large $\beta$, which we call an efficiency
parameter. Moreover, the entropy term leads to a distributed algorithm achieving the solution of {\bf A-CG-OPT} in following way. 

Regarding the parameter $\bm \theta$ as a dual variable of the
Lagrangian of {\bf A-CG-OPT}, its dual problem is simply represented
as $\min_{{\bm \theta}} \set{D}(\bm \theta)$ (see  for a detailed form of $\set{D}(\cdot)$). Then, the steepest
descent method to solve this dual problem with the direction
${\bm d}[t] = -\grad \set{D}({\bm \theta}[t])$ and step-size $a[t]$ is
given by
${\bm \theta}[t+1] = {\bm \theta}[t] + a[t] \cdot {\bm d}[t]$, \ie,
\begin{align} \label{eq:steep-dual}
\theta_i[t+1] &= \theta_i[t] + a[t] \Big( C_i'^{-1} \Big(\frac{-\theta_i[t]}{\beta}\Big) - s_i({\bm \theta}[t]) \Big), \cr
\theta_{ij}[t+1] &= \theta_{ij}[t] + a[t] \Big( U_{ij}'^{-1} \Big(\frac{\theta_{ij}[t]}{\beta}\Big) - s_{ij}({\bm \theta}[t]) \Big).
\end{align}

We highlight that {\bf \em Coord-dual} in \eqref{eq:Q-dual} is a
distributed implementation of \eqref{eq:steep-dual}, where the key
idea is to use (i) the instant rate $\hat{s}_i[t], \hat{s}_{ij}[t]$
from the current samples with (ii) diminishing step-size $a[t]$ (\ie,
satisfying \eqref{eq:dim-step}), instead of computing the exact rate
$s_i({\bm \theta}[t]), s_{ij}({\bm \theta}[t])$. Recall that computing
the service rate directly requires information of all other nodes and
edges, and measuring the service rate (\ie, the marginal probability
in the Markov chain induced by {\bf CDM}(${\bm \theta}[t]$)) requires
a {\em mixing time} to reach the stationary distribution from a large
number of samples. The proof of convergence and optimality of {\bf \em
  Coord-dual} using $\hat{\bm s}[t]$ with $a[t]$ is due to the
stochastic approximation technique \cite{borkar:SA, borkar:noise,
  PY10RA}, as presented in Section~\ref{sec:analysis}.

\smallskip
\noindent{\bf (b) {\em Coord-steep:}} 
Taking the different perspective of {\bf CG-OPT}, at any time, we sample the configuration via {\bf CDM}, offering the steepest ascent direction for $\set{F}(\bm \mu)$ in \eqref{eq:opt2}. Among feasible coordinates (\ie, elements) of ${\bm \mu}=[{\mu}_{\bm \sigma}]_{{\bm \sigma} \in \set{I}(G)}$, the steepest coordinate ascent method to solve \eqref{eq:opt2} deduces to select a configuration ${\bm \sigma}_\star$ according to the rule\footnote{A partial derivative of a function $f$ at the point
  ${\bm x}$ with respect to the $i$-th variable $x_i$ is denoted by
  $\frac{\partial f(\bm x)}{\partial x_i}$, or simply
  $\grad_{i}f(\bm x)$.}:
\begin{eqnarray*}
  {\bm \sigma}_\star= \arg \max_{{\bm \sigma} \in \set{I}(G)} \grad_{\bm \sigma} \set{F}({\bm \mu}),
\end{eqnarray*}
where from \eqref{eq:opt2}, 
\begin{eqnarray} \label{eq:steep-grad}
  \frac{\partial \set{F}(\bm \mu)}{\partial {\mu}_{\bm \sigma}} = \sum_{(i,j)\in E} \sigma_i\sigma_j  U_{ij}'( \mathbb{E}_{\bm \mu}[\sigma_i\sigma_j]) - \sum_{i \in V} \sigma_i C_i'( \mathbb{E}_{\bm \mu}[\sigma_i]).
\end{eqnarray}

Then, sampling configurations from the distribution that concentrates
on ${\bm \sigma}_\star$, \eg, a distribution (parameterized by
$\beta$), say $\bar{\bm \mu}$, such that $\bar{\mu}_{\bm \sigma} \propto \exp( \beta \cdot \grad_{\bm
  \sigma} \set{F}(\bar{\bm \mu}))$, approximates what the perfect
steepest ascent method would do. 
Therefore, from
\eqref{eq:ccd-stationary} and \eqref{eq:steep-grad}, the steepest
ascent method solving \eqref{eq:opt2} is approximated via
$\textbf{CDM}(\bm \theta)$ by setting $\bm \theta$ as follows:
\begin{align} \label{eq:steep-dir}
\theta_i = - \beta \cdot C_i'(\mathbb{E}_{\bm \theta}[\sigma_i]), ~\mbox{}~ \theta_{ij} = \beta \cdot U_{ij}'(\mathbb{E}_{\bm \theta}[\sigma_i \sigma_j]).
\end{align}

Note that the target parameter $\bm \theta$ is a fixed point of
\eqref{eq:steep-dir}, and its distribution $p_{\bm \theta}$ depends on
the marginal probability
$\mathbb{E}_{\bm \theta}[{\bm \phi}(\bm \sigma)]$, which may evolve
over time. Now, a fixed point iteration method of \eqref{eq:steep-dir}
is given by
\begin{eqnarray*}
\theta_i[t+1] = - \beta C_i'(\mathbb{E}_{{\bm \theta}[t]}[\sigma_i]), \ \theta_{ij}[t+1] = \beta U_{ij}'(\mathbb{E}_{{\bm \theta}[t]}[\sigma_i \sigma_j]).
\end{eqnarray*}
To smooth out the effect of random movements of the marginal
probability in \eqref{eq:steep-dir} and take a fixed point in a limit,
we consider an {\em exponential moving average} (EMA) with a constant
smoothing parameter $\alpha \in (0,1]$ as follows:
\begin{align} \label{eq:steep}
  \theta_i[t+1] &= \alpha \Big( -\beta C_i'(\mathbb{E}_{{\bm \theta}[t]}[\sigma_i])\Big) + (1-\alpha) \theta_i[t], \cr
   \theta_{ij}[t+1] &= \alpha \Big( \beta U_{ij}'(\mathbb{E}_{{\bm \theta}[t]}[\sigma_i\sigma_j])\Big) + (1 - \alpha)\theta_{ij}[t]. 
\end{align}

The key rationale of {\bf \em Coord-steep} in \eqref{eq:Q-steep}
towards a distributed operation of \eqref{eq:steep} is to use the
cumulative rate $\bar{\bm s}[t]$ instead of
$\mathbb{E}_{{\bm \theta}[t]}[{\bm \phi}(\bm \sigma)]$, which can
guarantee the convergence to the optimal point (for large $\beta$),
again, due to the stochastic approximation technique. In particular,
from \eqref{eq:agg-instant}, we have
\begin{eqnarray} \label{eq:serv_rate}
  \bar{\bm s}[t] = \bar{\bm s}[t-1] - \frac{1}{t}(\bar{\bm s}[t-1] -
  \hat{\bm s}[t]), \quad t \in \mathbb{Z}_{\geq 0},
\end{eqnarray}
thus the use of cumulative rate $\bar{\bm s}[t]$ has a similar effect
of exploiting instant rate $\hat{\bm s}[t]$ with $\frac{1}{t}$
step-size (\ie, satisfying \eqref{eq:dim-step}), as in {\bf \em
  Coord-dual}.

\smallskip
\noindent{\bf (c) {\em Coord-ind:}} 
A simple, myopic approach
  for a distributed operation to solve {\bf CG-OPT} would be to
  decompose its objective into node/edge-wise local optimization
  problems, \ie, node $i$ minimizes its cost and edge $(i,j)$
  maximizes its utility. Considering long-term activation/coordination
  rates under the stationary distribution $p_{\bm \theta}$, we
  associate each component of the parameter $\bm \theta$ with each
  local problem, \ie, $\min_{\theta_i} C_i(s_i(\bm \theta))$ and
$\max_{\theta_{ij}} U_{ij}(s_{ij}(\bm \theta))$. 
However, such a
myopic approach does not guarantee to achieve the optimal solution of
{\bf CG-OPT} due to the inter-coupling from $\bm \theta$ in the
objective functions. To reflect this inter-coupling among nodes and
edges, we design the following problem {\bf A-IND-OPT}\footnote{We
  denote the parameter vector for all other components except node $i$
  by $\theta_{-i}$, \ie, ${\bm \theta} = (\theta_i, \theta_{-i})$, and
  similarly $\theta_{-ij}$ for edge $(i,j)$.} with new objective
function, denoted by $\Psi_i(\theta_i)$ for node $i$ and
$\Psi_{ij}(\theta_{ij})$ for edge $(i,j)$, where the key part lies in
including {\em artificially-designed} penalty terms in
$\Psi_i(\theta_i)$ and $\Psi_{ij}(\theta_{ij})$.

\vspace{-0.35cm}
\separator \vspace{-0.4cm}
\begin{align} \label{eq:iaopt}
  \text{\bf (A-IND-OPT)} \quad 
  &\max_{\theta_i \in \mathbb{R}} ~\mbox{} \Psi_i(\theta_i), \quad i \in V, \cr
  &\max_{\theta_{ij} \in \mathbb{R}} ~\mbox{} \Psi_{ij}(\theta_{ij}), \quad (i,j) \in E, \cr
 \text{where} \quad \Psi_i(\theta_i) =& -C_i(s_i(\bm \theta)) - \frac{1}{\beta} \int_{-\infty}^{\theta_i} x s_i'(x,\theta_{-i}) \mathrm{d}x, \cr
  \Psi_{ij}(\theta_{ij}) = \mbox{} U_{ij}&(s_{ij}(\bm \theta)) - \frac{1}{\beta} \int_{-\infty}^{\theta_{ij}} x s_{ij}'(x,\theta_{-ij}) \mathrm{d}x.
\end{align}
\vspace{-0.25cm} \separator \vspace{-0.1cm}

The basic rationale of {\bf A-IND-OPT} is that each node and edge
chooses its own parameter by considering only its own cost or utility,
yet it might lead to the network-wide optimal status when imposing an
appropriate amount of penalty, \ie, the second term of
\eqref{eq:iaopt}. The form of penalty (parameterized by $\beta$) is of
critical importance to achieve the global optimal solution. This individually strategic form is well
understood by a game-theoretic perspective, as presented in
Section~\ref{sec:game_design}.

Now, the steepest ascent method to solve {\bf A-IND-OPT} with the
direction $d_i[t] = \grad \Psi_i(\theta_i[t])$,
$d_{ij}[t] = \grad \Psi_{ij}(\theta_{ij}[t])$ (see Appendix for the full derivation) and the step-size $a[t]$ is given by
${\bm \theta}[t+1] = {\bm \theta}[t] + a[t] \cdot {\bm d}[t]$, \ie,
\begin{align} \label{eq:steep-ind1}
  \theta_i[t+1] &= \theta_i[t] + \frac{a[t]}{\beta}\frac{\partial s_{i}({\bm \theta}[t])}{\partial \theta_{i}} \Big(-\beta C_i'(s_i({\bm \theta}[t])) - \theta_i[t] \Big), \cr
\theta_{ij}[t+1] &= \theta_{ij}[t] + \frac{a[t]}{\beta}\frac{\partial s_{ij}({\bm \theta}[t])}{\partial \theta_{ij}} \Big(\beta U_{ij}'(s_{ij}({\bm \theta}[t])) - \theta_{ij}[t] \Big).
\end{align}

Similarly to two earlier schemes, the key technique of {\bf \em
  Coord-ind} is a stochastic approximation, \ie, exploiting the
cumulative rate $\bar {\bm s}[t]$ from samples, instead of computing
the exact ${\bm s}({\bm \theta}[t]), \grad{\bm s}({\bm
  \theta}[t])$. The role of adopting the cumulative rate
$\bar{\bm s}[t]$ towards convergence to optimality can be clearly seen
by introducing the following alternative sequence, say
$\{{\bm \eta}[t]\}_{t \in \mathbb{Z}_{\geq 0}}$, which will be shown
to track the sequence
$\{{\bm \theta}[t]\}_{t \in \mathbb{Z}_{\geq 0}}$ of
\eqref{eq:steep-ind1} using $\bar{\bm s}[t]$ with a constant step-size
$a[t] = \alpha \in (0,1]$ (see Appendix for details),
defined as:
\begin{align} \label{eq:Q-aGD}
  \eta_i[t+1] = -\beta C_i'(\bar{s}_i[t]), \quad
  \eta_{ij}[t+1] = \beta U_{ij}'(\bar{s}_{ij}[t]).
\end{align}
From \eqref{eq:serv_rate}, the iterative update rule of the
alternative sequence is represented as follows: for large $t$,
\begin{eqnarray*} \label{eq:steep-alt}
\eta_i[t+1] = \eta_i[t] + \frac{1}{t} g_i(\eta_i[t]) \Big( C_i'^{-1} \Big(\frac{-\eta_i[t]}{\beta}\Big) - \hat{s}_i[t] \Big), \cr
\eta_{ij}[t+1] = \eta_{ij}[t] + \frac{1}{t} g_{ij}(\eta_{ij}[t]) \Big( U_{ij}'^{-1} \Big(\frac{\eta_{ij}[t]}{\beta}\Big) - \hat{s}_i[t] \Big),
\end{eqnarray*} 
where
\begin{eqnarray} \label{eq:g-fun}
  g_i(x) = \beta C_{i}''\Big(C_{i}'^{-1}\Big(\frac{-x}{\beta}\Big)\Big), ~\mbox{}
  g_{ij}(x) = -\beta U_{ij}''\Big(U_{ij}'^{-1}\Big(\frac{x}{\beta}\Big)\Big).
\end{eqnarray}
Note that the alternative process \eqref{eq:Q-aGD} has an effect of
exploiting $\hat{\bm s}[t]$ with diminishing step-size $\frac{1}{t}$,
as in {\bf \em Coord-dual}.

\subsection{Convergence and Optimality Analysis} \label{sec:analysis}

For provable convergence analysis, we first make the following
assumption, implying that we choose 
$\theta^{\text{min}}$ and $\theta^{\text{max}}$, such that the interval
$[\theta^{\text{min}}$, $\theta^{\text{max}}]$ is large enough to
include the optimal solution of {\bf A-CG-OPT}\footnote{The explicit
  values of $\theta^{\text{min}}$ and $\theta^{\text{max}}$ can be
  also computable as in \cite{jiang:distributed}.}.

\smallskip
\noindent {\bf (A1)} If ${\bm \theta}^0 \in \mathbb{R}^{|V|+|E|}$
solves for all $i \in V$ and $(i,j) \in E$,
\begin{align*}
  \theta_i^0 = -\beta C_i'\Big( \mathbb{E}_{{\bm \theta}^0}[\sigma_i]\Big), \quad 
  \theta_{ij}^0 = \beta U_{ij}'\Big( \mathbb{E}_{{\bm \theta}^0}[\sigma_i\sigma_j]\Big),
\end{align*}
then $\theta^{\text{min}} \leq \theta_i^0 \leq \theta^{\text{max}}$
and $\theta^{\text{min}} \leq \theta_{ij}^0 \leq
\theta^{\text{max}}$. Note that, for example, if
$U_{ij}'(0) < \infty$, then {\bf (A1)} for $\theta_{ij}^0$ is
satisfied when $\theta^{\text{min}} \leq \beta U_{ij}'(1)$ and
$\theta^{\text{max}} \geq \beta U_{ij}'(0)$.

\smallskip
Now, the next theorem is our main result, which states the convergence
of {\bf \em Coord}-algorithms to a point arbitrarily close to the
optimal solution of {\bf CG-OPT}, under some mild conditions.

\vspace{-0.35cm}
\separator
\vspace{-0.35cm}
\begin{theorem}[Convergence/Optimality of {\bf \em Coord}-algorithms]
  \label{thm:conv} \mbox{}
\begin{compactenum}[(i)]
\smallskip
\item \underline{\em Convergence.} Under {\bf (A1)}, for strictly
  concave/convex, continuously twice-differentiable utility/cost functions,  choose a
  step-size function $a[\cdot]$ in {\bf \em Coord-dual} satisfying
  \begin{eqnarray} \label{eq:dim-step}
    \sum_t a[t] = \infty, \quad \sum_t a[t]^2 < \infty.
  \end{eqnarray}
  Then, for any initial condition ${\bm \theta}[0]$, under all {\bf
    \em Coord}-algorithms, ${\bm \theta}[t]$ and corresponding
  $\bar{\bm s}[t]$ (from \eqref{eq:agg-instant}) converges to
  $({\bm \theta}^\circ, {\bm \lambda}^\circ)$, \ie,
\begin{eqnarray*}
  \lim_{t \rightarrow \infty} {\bm \theta}[t] = {\bm \theta}^\circ ~\text{ and } \lim_{t \rightarrow \infty} {\bar{\bm s}}[t] = {\bm \lambda}^\circ, ~\text{ almost surely,}
\end{eqnarray*}
where $({\bm \theta}^\circ, {\bm \lambda}^\circ)$ is such that
$(p_{{\bm \theta}^\circ}, {\bm \lambda}^\circ)$ attains the (unique)
solution of \textbf{A-CG-OPT} in \eqref{eq:aopt} (over $\bm \mu$ and
$\bm \lambda$).

\smallskip
\item \underline{\em Optimality.} Furthermore, {\bf \em
    Coord}-algorithms approximately solve \textbf{CG-OPT} in the
  following sense:
  \begin{align} \label{eq:beta}
& \sum_{(i,j) \in E} U_{ij}(\lambda_{ij}^\circ) - \sum_{i \in V} C_i(\lambda_i^\circ) \geq \cr
& \quad \sum_{(i,j) \in E} U_{ij}(\lambda_{ij}^\star) - \sum_{i \in V} C_i(\lambda_i^\star) - \frac{\log |\set{I}(G)|}{\beta},
  \end{align}
  where ${\bm \lambda}^\star$ is the optimal solution of {\bf CG-OPT}
  in \eqref{eq:opt}.
\end{compactenum}
\end{theorem}
\vspace{-0.45cm}
\separator
\vspace{-0.15cm}

The proof of Theorem~\ref{thm:conv} is presented in Appendix, but we briefly provide the proof sketch for readers'
convenience. Each scheme of {\bf \em Coord}-algorithms is interpreted
as a stochastic approximation procedure with controlled Markov noise,
and a main technical challenge lies in handling a non-trivial coupling
between Markov process of $\textbf{CDM}(\bm \theta)$ and parameter
$\bm \theta$ updates. Simply, a provable convergence is guaranteed on
the strength of stochastic approximation theory, in that we intuitively expect that by exploiting (i) instant rate
$\hat{\bm s}[\cdot]$ with diminishing step-size in \eqref{eq:dim-step}
or (ii) cumulative rate $\bar{\bm s}[\cdot]$ which has an effect of
$\frac{1}{t}$ step-size (\ie, satisfying \eqref{eq:dim-step}), the speed of variations of the parameter $\bm \theta$ tends to zero
after sufficiently long time. Thus, its limiting behavior can be
understood by ordinary differential equations (ODE). We highlight that
we adopt diminishing step-size in \eqref{eq:dim-step}, following the
standard ODE approaches in stochastic approximation theory as in
\cite{borkar:SA, wasan:SA, kushner:SA} and references therein, to
provide provable convergence. The additional challenge dealing with
{\bf \em Coord-steep} and {\bf \em Coord-ind} (not existing for {\bf
  \em Coord-dual}) is that they have higher-order temporal
dependencies in their updating rules, \ie, use the current parameter
${\bm \theta}[t]$ directly when obtaining the next parameter
${\bm \theta}[t+1]$. To handle this issue, we define `alternative'
process (see $\{{\bm \rho}[t]\}_{t \in \mathbb{Z}_{\geq 0}}$ and
$\{{\bm \eta}[t]\}_{t \in \mathbb{Z}_{\geq 0}}$ in Appendix) and argue its convergence under the relation to that of the
original process $\{{\bm \theta}[t]\}_{t \in \mathbb{Z}_{\geq 0}}$.


%% file: approaches.tex
\section{New Interpretations via Game Theory} \label{sec:interpretation}

In Section~\ref{sec:alg_analysis}, we develop three simulation-based
algorithms that adaptively update the parameter $\bm \theta$ in a
distributed manner, but result in the optimal solution of {\bf
  CG-OPT}.  The rationale behind each scheme follows the framework of
distributed optimization.  In this section, we take a different angle
to reverse engineer two algorithms {\bf \em Coord-steep} and {\bf \em
  Coord-ind} in other framework, which is the game-theoretic one.  As
a background, game theory has been emerged as a powerful tool not only
to analyze the rational behavior of competitive multi-agent systems
(\ie, just optimizing a local objective), but also to control the
local behavior of each agent, see \eg, \cite{NJ13}. In such a
framework, it is aimed that a game is designed with an {\em
  artificially-selected} payoff function so that local decisions of
agents result in a system-wide desirable solution such as an unique,
fair or socially-optimal point. Moreover, a game-theoretic approach
provides valuable insights into the design of various robust local
control rules through (distributed) game dynamics, whereas the
standard centralized optimization framework can not directly consider
the interactions among agents. In this section, inheriting such a philosophy of the game-theoretic
framework for distributed optimization, we establish the desirable
properties of equilibrium from a well-designed non-cooperative game
and present that {\bf \em Coord-steep} and {\bf \em Coord-ind}
correspond to the stochastically-approximated variants of two popular
game-learning dynamics.

\subsection{$\text{CoordGain}(\beta)$ Game} \label{sec:game_design}

We first design a non-cooperative game, denoted by
$\text{CoordGain}(\beta)$ with $\beta >0$.

\vspace{-0.3cm}
\separator
\vspace{-0.2cm}
$\text{CoordGain}(\beta)$
\vspace{-0.4cm}
\separator
\vspace{-0.2cm}
\begin{compactenum}[(i)]
\item {\bf \em Players. }
  Each node $i \in V$ and each edge $(i,j) \in E$ acts as a
  player. Let $N= V \cup E$ denote the set of players, and thus
  $n \in N$ can be either a node $i \in V$ or an edge
  $(i,j) \in E$\footnote{We interchangeably use a notation of $ij$ and
    $(i,j)$ for an edge player.}.
  
\item {\bf \em Strategy. } Each player $n$ has a parameter
  $\theta_n \in \mathbb{R}$ as its own strategy. We denote the
  strategy profile of entire players by
  ${\bm \theta} = [\theta_n]_{n \in N} = ([\theta_i]_{i \in V},
  [\theta_{ij}]_{(i,j) \in E}) \in \mathbb{R}^{|N|}$.
  
\item {\bf \em Payoff. } The payoff function of player $n \in N$,
  denoted by
  $\Psi_n(\theta_n, \theta_{-n}): \mathbb{R}^{|N|} \mapsto
  \mathbb{R}$, is designed to be a net-coordination utility (or
  net-activation cost) with incurring {\em penalty} function
  $V_n(\cdot)$ resulting from coordination:
  \begin{align} \label{eq:payoff}
    \Psi_i(\theta_i,  \theta_{-i}) &= -C_i(s_i({\bm \theta})) -
    \frac{1}{\beta} V_i (\theta_i, \theta_{-i}), \cr
    \Psi_{ij}(\theta_{ij},  \theta_{-ij}) &= U_{ij}(s_{ij}({\bm \theta}))
    - \frac{1}{\beta} V_{ij} (\theta_{ij}, \theta_{-ij}), 
  \end{align}
 \vspace{-0.3cm}
  \begin{align}
    \label{eq:penalty}
~\text{where} \quad  V_i(\theta_i, \theta_{-i}) &= \int_{-\infty}^{\theta_i} x
    s_i'(x,\theta_{-i}) \mathrm{d}x, \cr
     V_{ij}(\theta_{ij}, \theta_{-ij}) &= \int_{-\infty}^{\theta_{ij}} x
    s_{ij}'(x,\theta_{-ij}) \mathrm{d}x.
  \end{align}
\end{compactenum}
\vspace{-0.2cm}
\separator
\vspace{-0.1cm}

Note that a player $n$'s payoff $\Psi_n(\cdot)$\footnote{We use the
  notation $\Psi_i, \Psi_{ij}$ in both \eqref{eq:iaopt} and
  \eqref{eq:payoff} for notational simplicity, since they obviously
  have the same detailed form.} is determined by how aggressively
other players are activated/coordinated as well as how itself
does. The parameter $\beta$ quantifies {\em penalty level} in the
players' payoffs, and we realize that it balances the trade-off
between the quality of equilibria and the convergence speed to the
equilibria under game dynamics (see Theorem~\ref{thm:NE}).

To achieve our goal of obtaining good equilibria and a provable
transfer to distributed game dynamics converging to an equilibrium,
the choice of penalty function $V_n(\cdot)$ is of crucial
importance. Our choice of penalty function \eqref{eq:penalty} captures
following two features.

First, it appropriately measures each player's
impact of excessive strategy on other players. One na{\"i}ve choice of
penalty to be imposed by a player $n$ may be
$V_n(\bm \theta) = \theta_n \times s_n(\theta_n, \theta_{-n})$, which
is proportional to the current strategy $\theta_n$ multiplied by its
achieved long-term gain $s_n(\bm \theta)$, yet it is unclear that this
penalty provides a provable framework of equilibrium analysis. On the
other hand, our design of penalty function considers the expected
strategy value $\mathbb{E}[\Theta_n]$ which depends on the {\em
  relative increasing speed} of one's rate in the interval
$(-\infty, \theta_n)$, by letting $\Theta_n \in [-\infty, \theta_n]$
denote a continuous random variable with the density function
$f_{\Theta_n}(x) = \frac{1}{ s_{n}(\theta_{n},\theta_{-n})}
\frac{\partial s_{n}(x,\theta_{-n})}{\partial x}$ so that the penalty
function is represented as
$V_n(\bm \theta) = \mathbb{E}[\Theta_n] \times s_n(\theta_n,
\theta_{-n})$.

Second, the penalty function \eqref{eq:penalty} is a
function of self-strategy and its marginal distribution, not the
individual strategy values or payoffs of others. From simple algebra
of \eqref{eq:marginal}, it is shown to be structured in terms of local
information:
$V_n({\bm \theta}) = \theta_n s_n(\bm \theta) + \ln (1-s_n(\bm
\theta))$. Since $s_n(\cdot)$ can be measured in the midst of playing
a player's own strategy, \eg, $\hat{s}_n(\cdot)$, via {\bf CDM} with
one-hop message passing, best response or payoff gradient of our game
can be locally estimated. This feature enables us to develop
distributed game dynamics, which indeed corresponds to {\bf \em
  Coord-steep} and {\bf \em Coord-ind} (see
Section~\ref{sec:game_dynamics}).

\subsection{Equilibrium Analysis} \label{sec:game_eq}

We first present popular notions: Nash equilibrium and
Price-of-Anarchy in game theory:
\begin{definition}
\label{def:NE} 
In the $\text{CoordGain}(\beta)$,  \\
(i) a strategy profile $\bm{\theta}^{\text{NE}}$ is a \emph{Nash
  equilibrium} (NE) if
\begin{align*}
  \Psi_n(\theta_n^{\text{NE}}, \theta_{-n}^{\text{NE}}) &\geq
  \Psi_n(\theta_n, \theta_{-n}^{\text{NE}}), \quad \forall \theta_n \in
  \mathbb{R}, ~ \forall n \in N.
\end{align*}
(ii) a {\em Price-of-Anarchy} (PoA) is
\begin{align*}
  \text{PoA} = \frac{ \max_{{\bm \theta}} \sum_{(i,j) \in
      E}U_{ij}(s_{ij}(\bm \theta)) - \sum_{i \in V} C_i(s_i(\bm \theta))
  }{ \min_{{\bm \theta}^{\text{NE}}} \sum_{(i,j) \in
      E}U_{ij}(s_{ij}(\bm \theta)) - \sum_{i \in V} C_i(s_i(\bm \theta))
  }.
\end{align*}
\end{definition}
Furthermore, we say that a NE $\bm{\theta}^{\text{NE}}$ (if exists) of
the game is {\em non-trivial}, if players' activation/coordination
rate at equilibrium ${\bm s}(\bm{\theta}^{\text{NE}})$ is positive,
and {\em trivial} otherwise. The PoA indicates the ratio between the
social optimum and the worst equilibrium of the game, and we say {\em
  no PoA} if $\text{PoA} = 1$. We now present our main results on the
equilibrium analysis.

\vspace{-0.35cm}
\separator
\vspace{-0.4cm}
\begin{theorem}[Uniqueness and PoA] \label{thm:NE} 
  In the $\text{CoordGain}(\beta)$, 
\begin{compactenum}[(i)]
\smallskip
\item \underline{\em Uniqueness.} for any $\beta >0,$ there exists a
  unique non-trivial NE ${\bm \theta}^{\text{NE}}.$
\smallskip
\item \underline{\em Price-of-Anarchy.}
  $(p_{{\bm \theta}^{\text{NE}}}, {\bm s}({\bm \theta}^{\text{NE}}))$
  attains the optimal solution of {\bf A-CG-OPT}, and thus
  $\lim_{\beta \rightarrow \infty} \text{PoA} = 1.$ 
\end{compactenum}
\end{theorem}
\vspace{-0.45cm}
\separator
\vspace{-0.15cm}

The proof of Theorem~\ref{thm:NE} is presented in Appendix. We prove that our game is an {\em ordinal potential game},
where the potential function corresponds to the dual function of {\bf
  A-CG-OPT}. It implies that our game has a unique non-trivial NE
${\bm \theta}^{\text{NE}}$, in particular, the solution of {\bf
  A-CG-OPT} is attained at
$(p_{{\bm \theta}^{\text{NE}}}, {\bm s}({\bm \theta}^{\text{NE}}))$,
\ie, ${\bm \theta}^{\text{NE}} = {\bm \theta}^o$. Therefore, there is
asymptotically no PoA in our game, \ie, the aggregate coordination
gain at the unique non-trivial NE becomes arbitrarily close to the
social optimum by choosing sufficiently large $\beta$.

\subsection{Reverse Engineering of {\bf \em Coord-steep} and {\bf \em
    Coord-ind}} \label{sec:game_dynamics}

\noindent{\bf \em Best response dynamics.} The most popular dynamics
is the {\em best response (BR) dynamics} that each player chooses its
best strategy, given the strategy (at the previous frame) of all other
players, \ie, at frame $t$,
\begin{align*}
  \theta_n[t+1] = \text{BR}_n(\theta_{-n}[t]) := \arg \max_{\theta_n \in
    \mathbb{R}} \Psi_n(\theta_n, \theta_{-n}[t]), 
\end{align*}
which leads to a fixed point of the following function in
$\text{CoordGain}(\beta)$: $\forall i \in V, \forall (i,j)\in E$,
\begin{eqnarray} \label{eq:BR}
\theta_i[t+1] &=& -\beta C_i'\Big( s_i(\theta_i[t+1], \theta_{-i}[t])
\Big), \cr
\theta_{ij}[t+1] &=& \beta U_{ij}'\Big( s_{ij}(\theta_{ij}[t+1], \theta_{-ij}[t]) \Big).
\end{eqnarray}

\smallskip
\noindent{\bf \em Jacobi dynamics.} The second dynamics is {\em Jacobi
  dynamics}, whose idea is to adjust each player's strategy gradually
towards its best response strategy, \ie, at frame $t$,
\begin{align*} 
  \theta_n[t+1] = \theta_n[t] +
  \alpha \cdot \big(\text{BR}_n(\theta_{-n}[t]) - \theta_n[t]\big),
\end{align*}
where $\alpha \in (0,1]$ is a smoothing parameter\footnote{Jacobi
  dynamics generally makes a smoother move than the BR dynamics, where
  a small smoothing parameter plays the role of compensating for the
  instability of the BR dynamics, see \cite{chen2007game}.}. The
smoothing parameter captures how accurately the dynamics follows the
BR dynamics, where $\alpha = 1$ corresponds to the BR dynamics. From
\eqref{eq:steep} and \eqref{eq:BR}, we can verify that a variant of
Jacobi dynamics of our game (BR dynamics as a special case) which
exploits cumulative rate $\bar{\bm s}[t]$ instead of
${\bm s}({\bm \theta}[t])$ at every frame is indeed equivalent to {\bf
  \em Coord-steep}, \ie, approximated (parameterized by $\beta$)
steepest ascent method of {\bf CG-OPT}.

\smallskip
\noindent{\bf \em Gradient dynamics.} Finally, we investigate the {\em
  gradient dynamics} \cite{flam02} that each player $n$ first
determines the gradient of its payoff \eqref{eq:payoff},
$\grad \Psi_n(\bm \theta)$, then updates its strategy in that
direction with a constant step-size $\alpha \in (0,1]$, \ie, at frame
$t$,
\begin{eqnarray*}
\theta_n[t+1] &=& \theta_n[t] + \alpha \cdot \nabla \Psi_n(\theta_n[t]).
\end{eqnarray*}
The interpretation of the gradient dynamics from an economic
perspective is that if the marginal coordination utility of an edge
$(i,j)$ exceeds the marginal penalty, \ie,
$\grad \Psi_{ij}(\bm \theta) > 0$, its strategy is increased to
achieve more coordination gain, and if
$\grad \Psi_{ij}(\bm \theta) < 0$, its strategy is decreased to reduce
the penalty. From the objective function \eqref{eq:iaopt} and our
payoff function \eqref{eq:payoff}, we can verify that {\bf \em
  Coord-ind} is equivalent to a variant of gradient dynamics of our
game, which exploits $\bar{\bm s}[t]$ from samples instead of
computing the exact
${\bm s}({\bm \theta}[t]), \grad {\bm s}({\bm \theta}[t])$.

\smallskip To summarize, {\bf \em Coord-steep}, {\bf \em Coord-ind}
are stochastically-approximated variants of Jacobi dynamics and
gradient dynamics of $\text{CoordGain}(\beta)$,
respectively. Theorem~\ref{thm:conv} states that those dynamics
converge to the unique non-trivial NE. Note that this is a new feature
in our work, not prevalent in general game-theoretic approaches for
distributed optimization, \ie, there exists no generalized distributed
dynamics converging to a NE (even it exists) due to a lack of
information, in a broad class of games \cite{hartcol03}.


%% file: simulations.tex
\section{Numerical Results}
\label{sec:sim}

In this section, we carry out numerical experiments to assess our
analytical findings of {\bf \em Coord}-algorithms by considering
networks with various topologies and cost functions.

\begin{figure}[t!]
  \centering 
  \subfigure[\small{{\tt STAR}}] {
    \includegraphics[width=0.17\columnwidth]{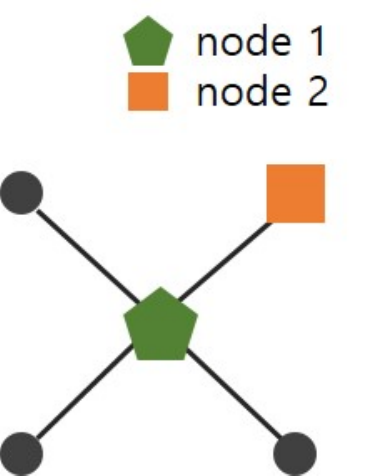}
    \label{fig:star_topology}} 
  \hspace{0.1cm}
  \subfigure[\small{{\tt COMP}}] {
    \includegraphics[width=0.15\columnwidth]{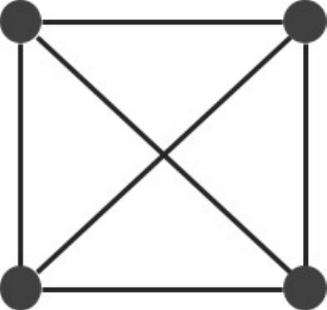}
    \label{fig:comp_topology}} 
  \hspace{0.1cm}  
  \subfigure[\small{{\tt RAND} topology}]
  { 
    \includegraphics[width=0.4\columnwidth]{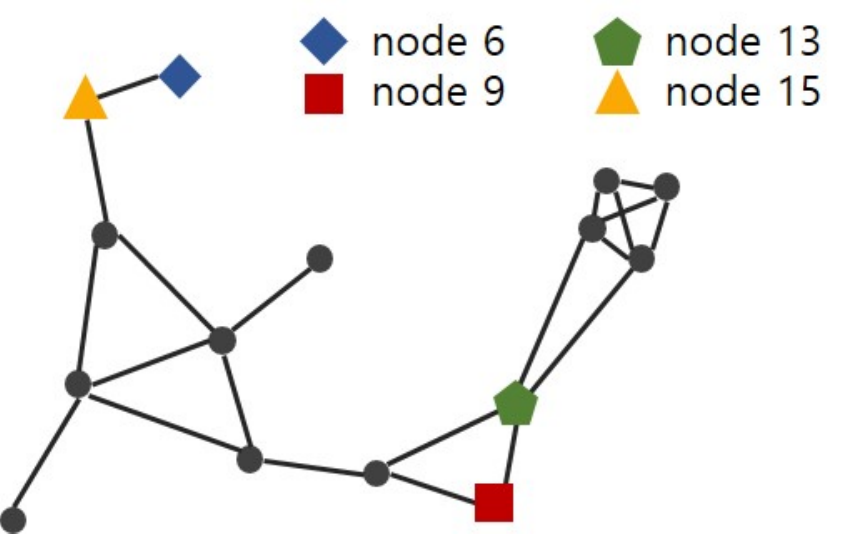}
    \label{fig:rg_topology}}   
  \caption{{Network topologies}}
  \label{fig:topology}
\end{figure}

\begin{figure*}[t!]
  \centering 
  \subfigure[\small{Parameter of {\tt STAR-C1}}]
  { 
    \includegraphics[width=0.4\columnwidth]{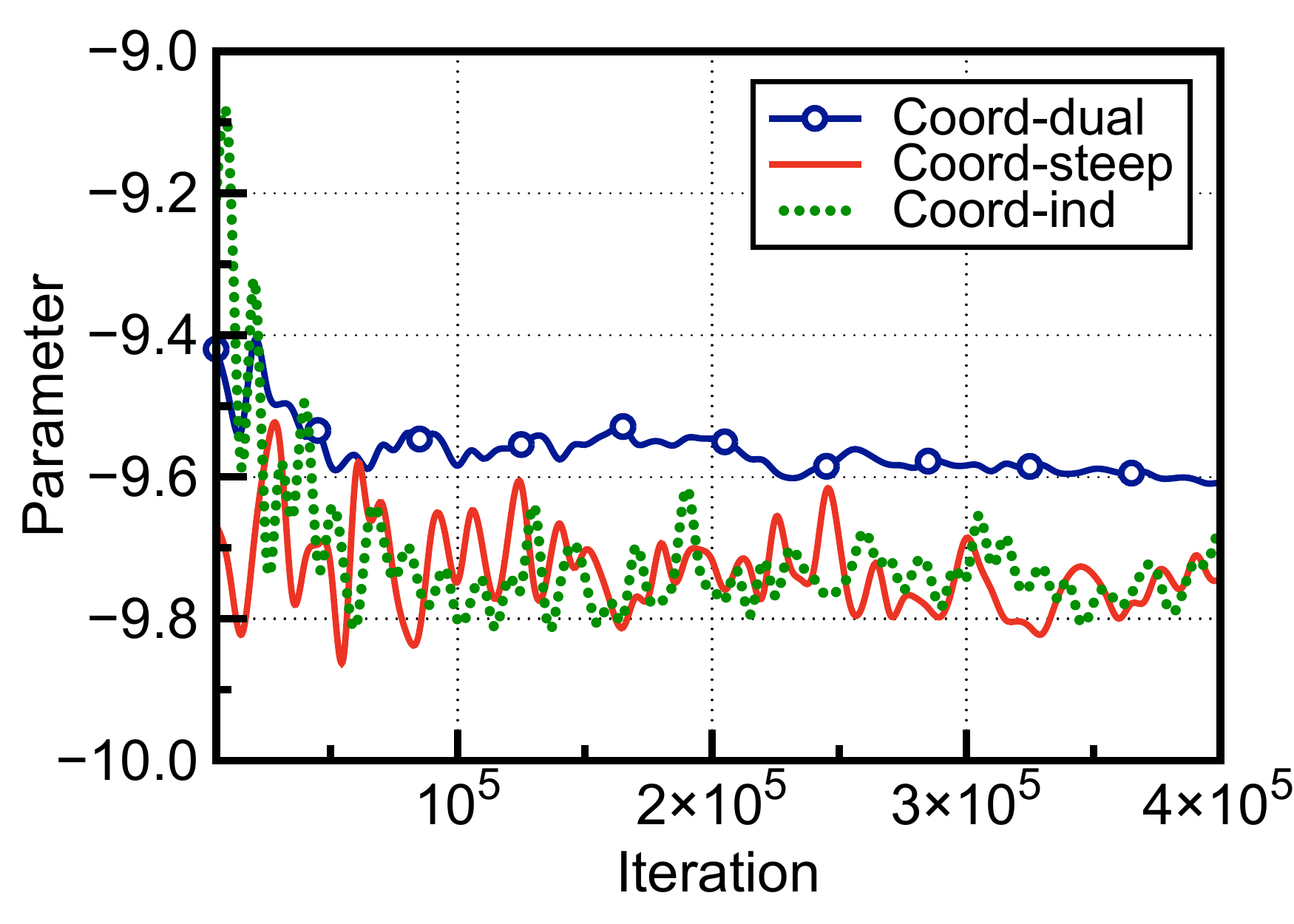}
    \label{fig:star-s1-strategy}} 
  \subfigure[\small{Coordination gain of {\tt STAR-C1}}]
  { 
    \includegraphics[width=0.4\columnwidth]{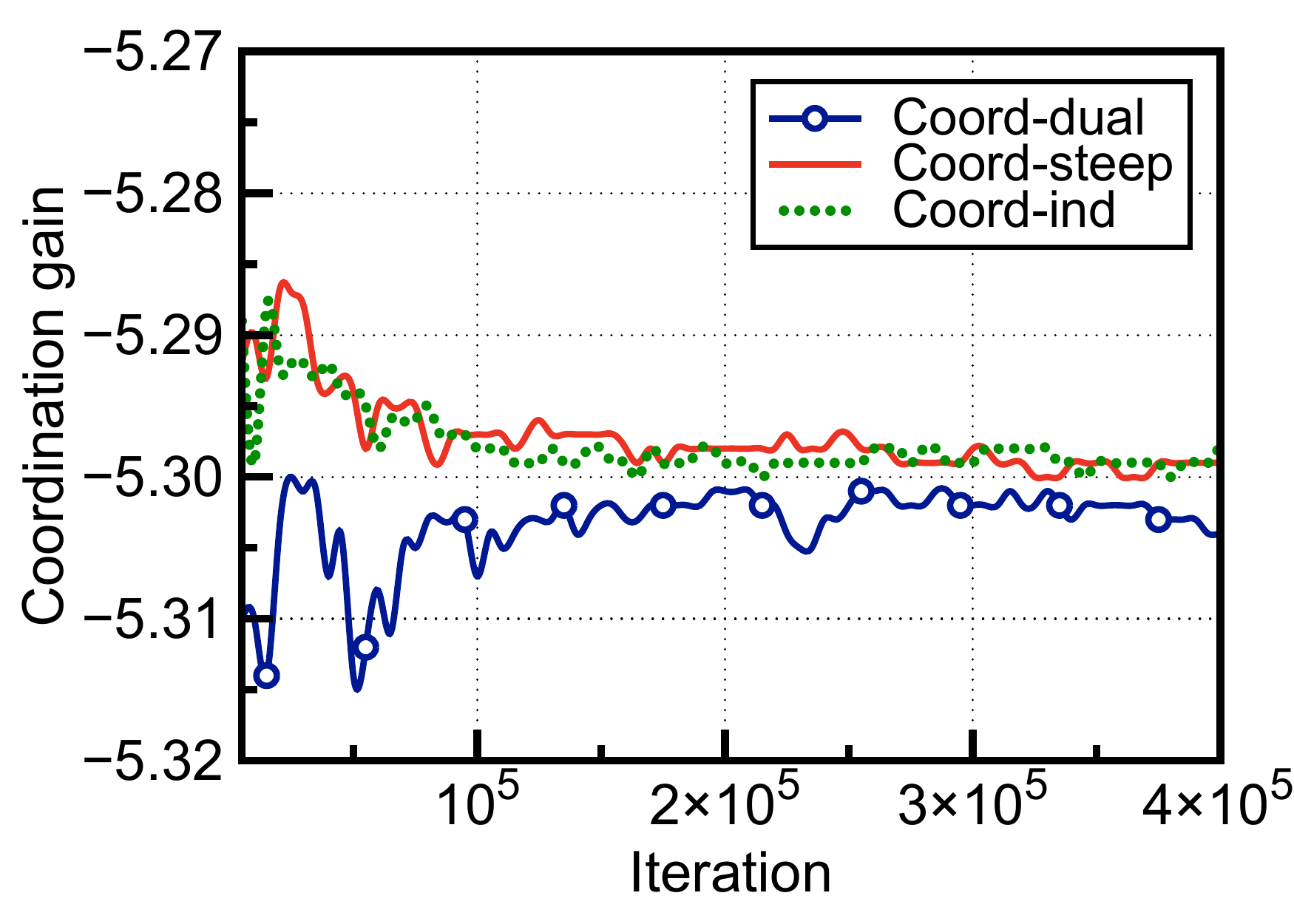}
    \label{fig:star-s1-profit}}   
  \subfigure[\small{Long-term rate of {\tt COMP-C1}}]
  { 
    \includegraphics[width=0.4\columnwidth]{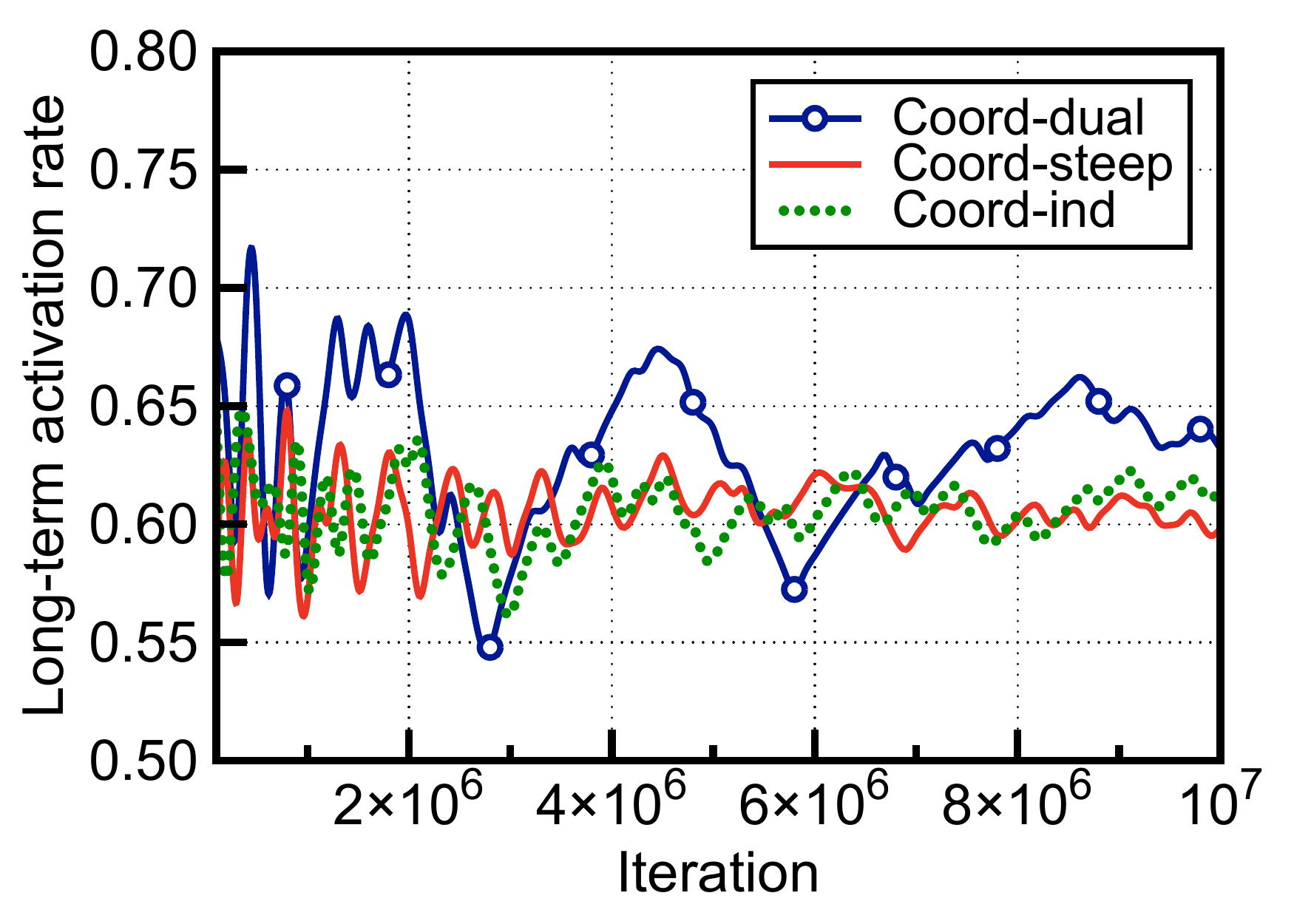}
    \label{fig:comp-s1-rate}} 
  \subfigure[\small{Coordination gain of {\tt COMP-C1}}]
  { 
    \includegraphics[width=0.4\columnwidth]{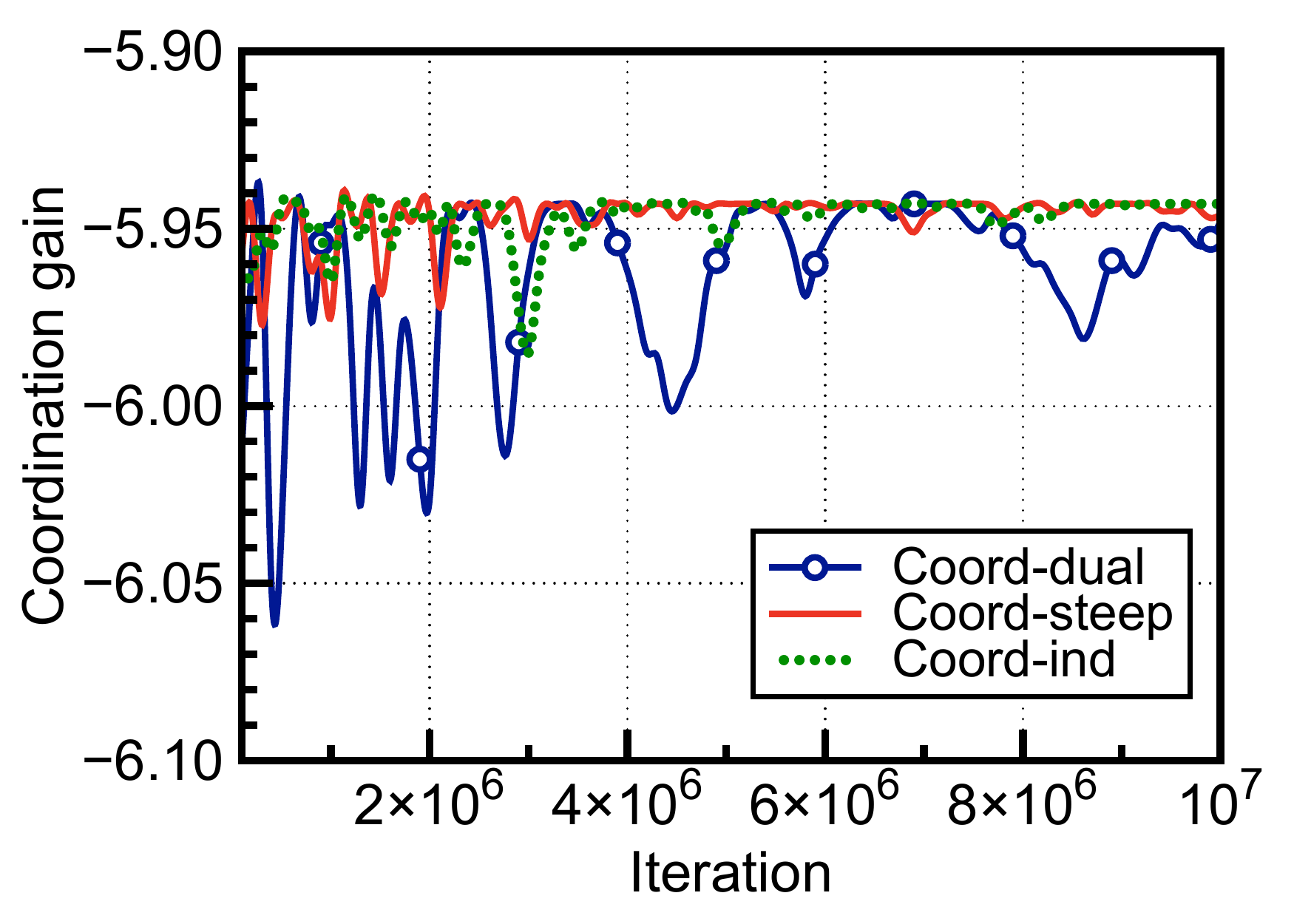}
    \label{fig:comp-s1-profit}} 
  \caption{{Convergence of parameter, coordination gain, and
      long-term rate to optimality on {\tt STAR-C1} and {\tt
        COMP-C1}.}}
  \label{fig:conv_simple}
\end{figure*}

\begin{figure*}[t!]
  \centering 
  \subfigure[\small{Coordination gain of {\tt RAND-C1}}] {
    \includegraphics[width=0.4\columnwidth]{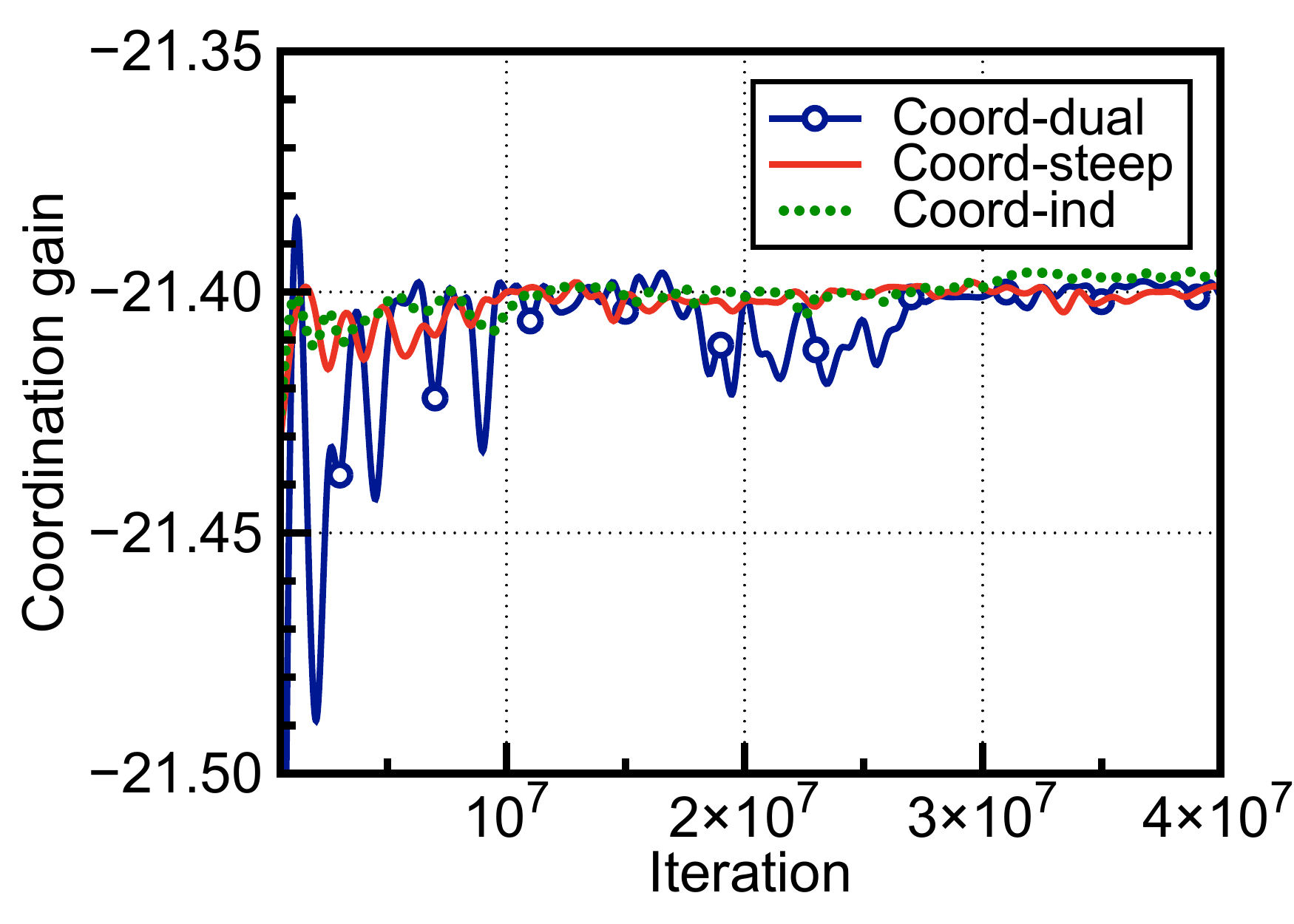}
    \label{fig:rand-s1-profit-comp}} 
\subfigure[\small{Long-term rate of {\tt RAND-C1}}]
  { 
    \includegraphics[width=0.4\columnwidth]{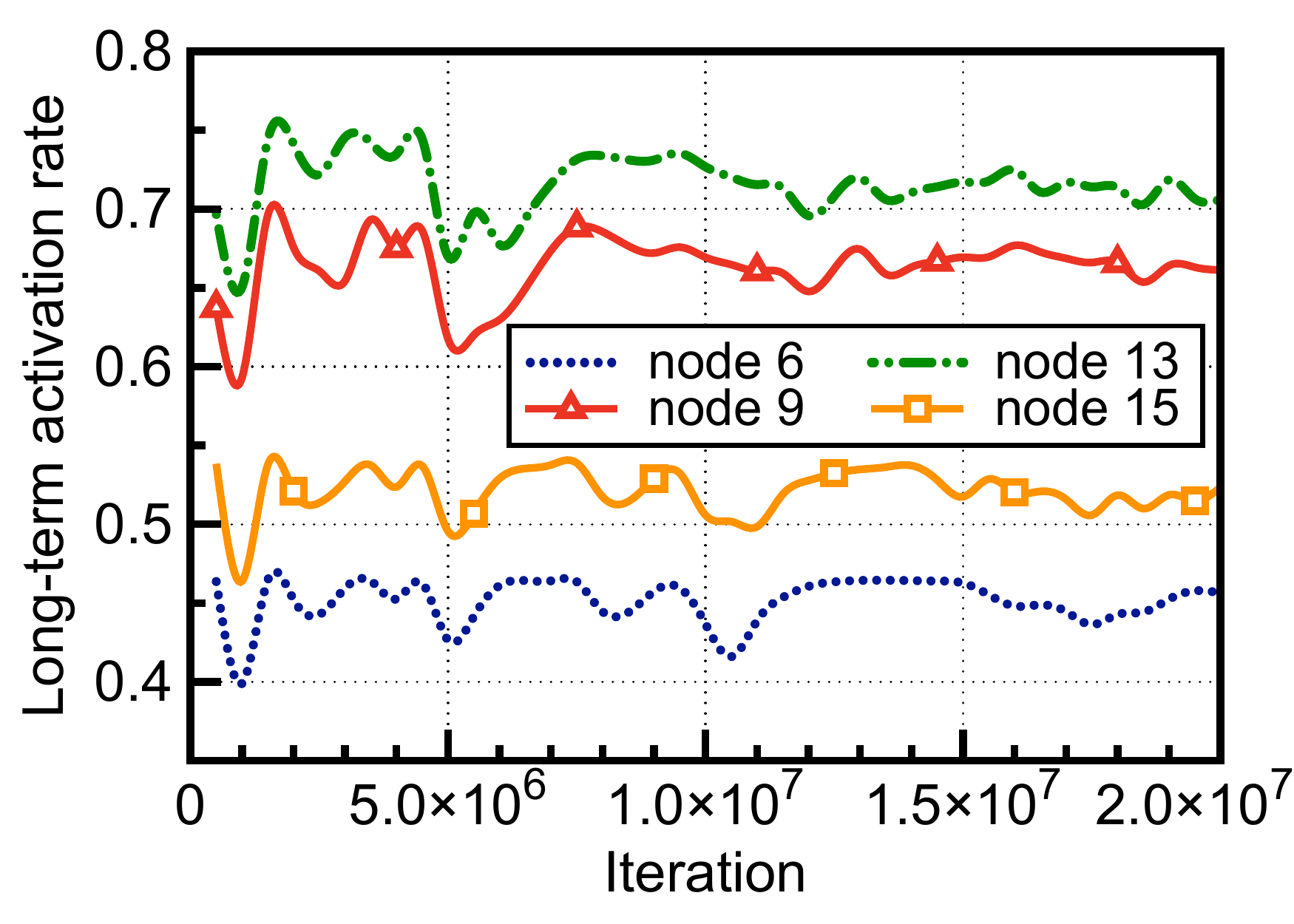}
    \label{fig:rand-s1-rate}} 
  \subfigure[\small{Long-term rate of {\tt RAND-C2}}]
  { 
    \includegraphics[width=0.4\columnwidth]{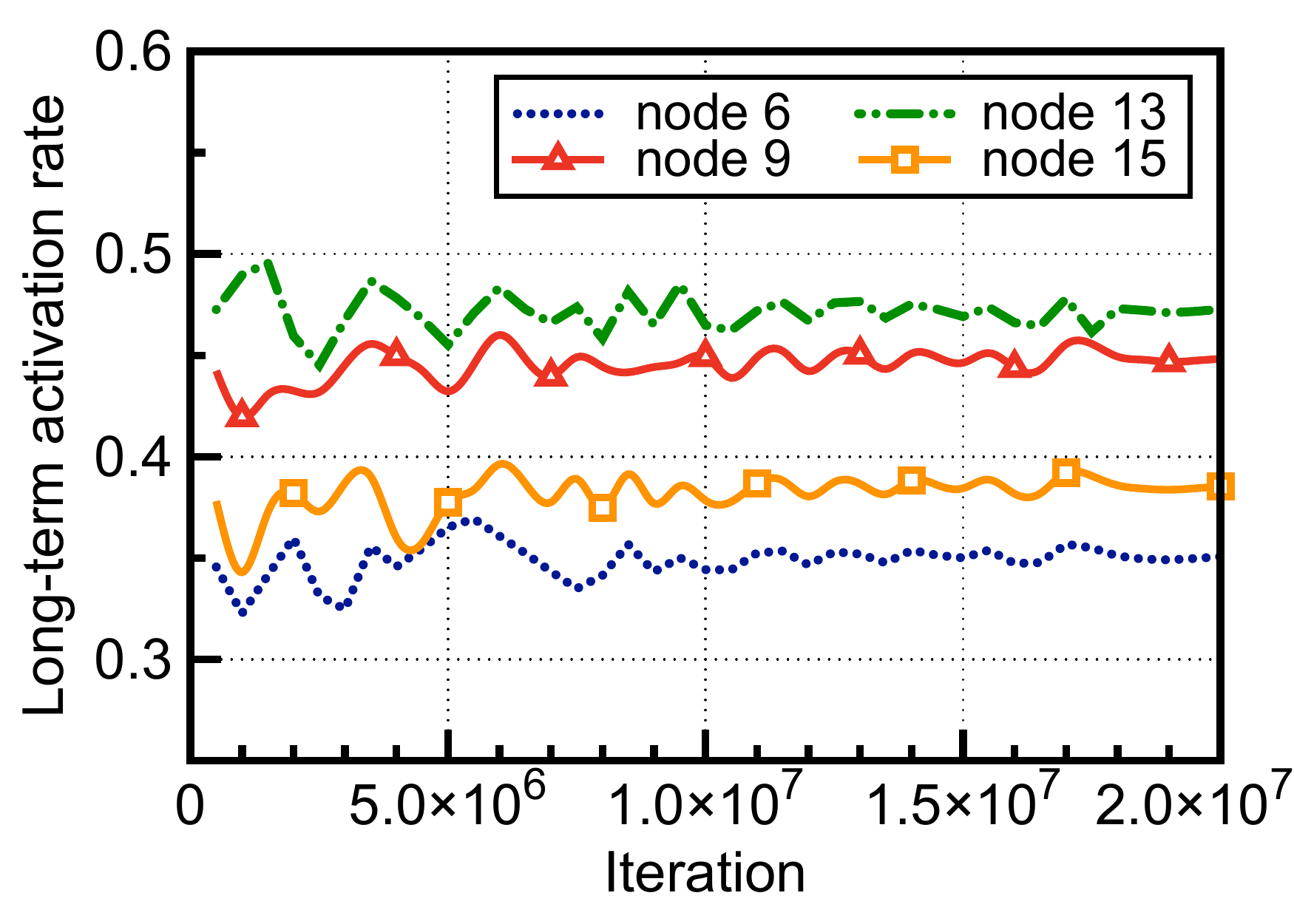}
    \label{fig:rand-s2-rate}} 
  \subfigure[\small{Trade-off between efficiency($\beta$) and
    convergence speed}]
  { 
    \includegraphics[width=0.4\columnwidth]{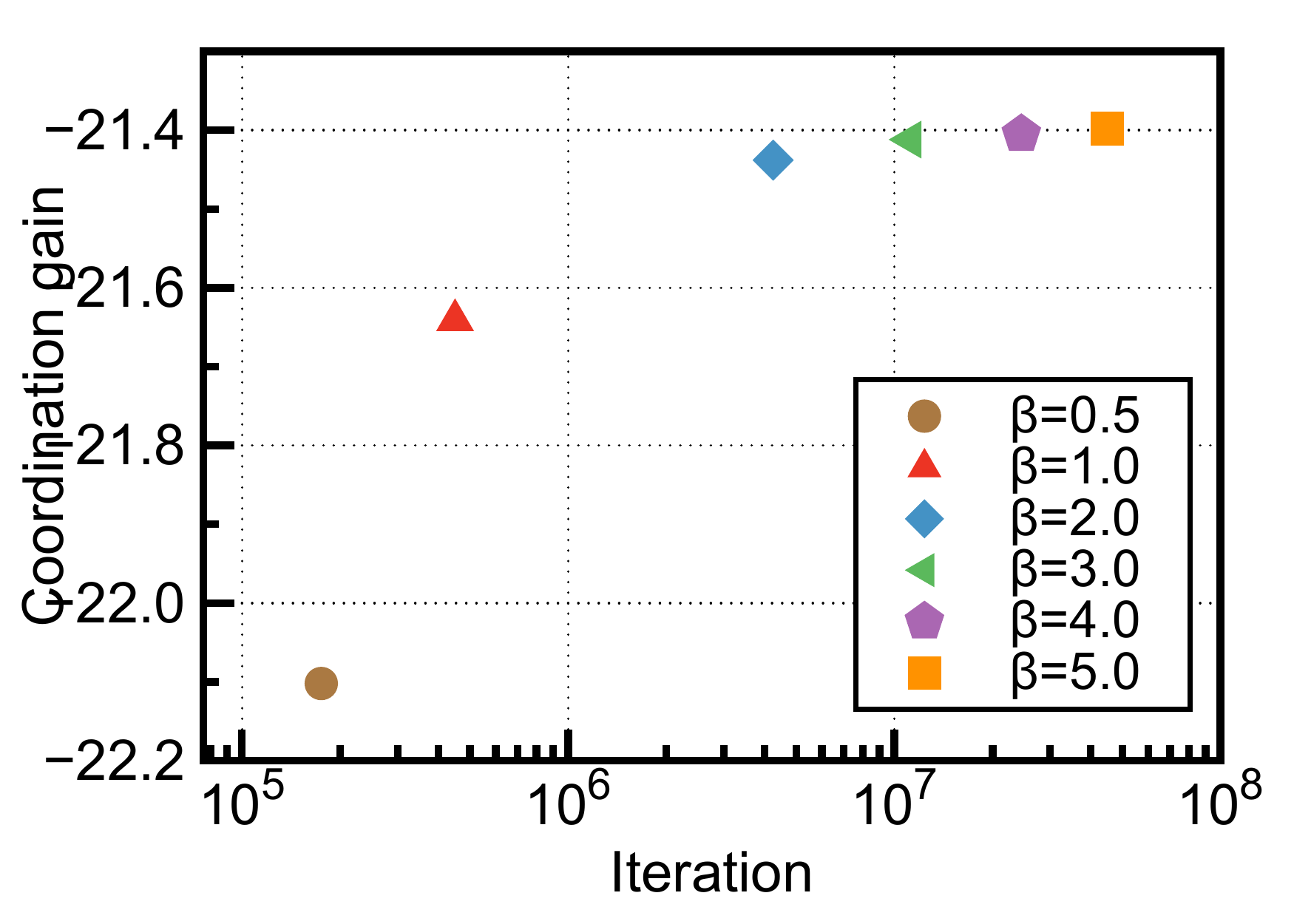}
    \label{fig:poa}} 
  \caption{{Convergence of coordination gain, long-term rate, and
      trade-off on {\tt RAND-C1} and {\tt RAND-C2}.}}
  \label{fig:conv_rand}
\end{figure*}

\smallskip
\noindent {\bf \em Setup.} In this paper, we consider ``basic''
topologies to show that our {\bf \em Coord}-algorithms converge to the
accurate solution, and a random topology that is regarded as a
collection of such basic topologies for more general results. The
network topologies under which our results are presented here are
star, complete, and random graphs. For numerical results, we consider
proportional fairness across edges for coordination utility, \ie,
$U_{ij}(x) = \log(x)$ for all edges $(i,j) \in E$, and consider two
cost functions for nodes: {\tt (C1)} $C_i(x) = 2x^2$ and {\tt (C2)}
$C_i(x) = \frac{1}{1-x}$ for all nodes $i \in V$, as classified into
the following $4$ topologies.

\begin{compactenum}[$\circ$]
\item {\tt STAR-C1}: Star graph of $5$ nodes with {\tt
    (C1)}
\item {\tt COMP-C1}: Complete graph of $4$ nodes with {\tt
    (C1)} 
\item {\tt RAND-C1}: Random graph of $15$ nodes, $21$ edges with {\tt
    (C1)}
\item {\tt RAND-C2}: Random graph of $15$ nodes, $21$ edges with {\tt
    (C2)}
\end{compactenum}

The above topologies are depicted in Fig.~\ref{fig:topology}:
Fig.~\ref{fig:star_topology} for {\tt STAR-C1}, Fig.~\ref{fig:comp_topology} for {\tt COMP-C1}, and
Fig.~\ref{fig:rg_topology} for {\tt RAND-C1}, {\tt RAND-C2}. Moreover,
for a fixed frame duration $T=10$, we choose a step-size function
$a[t] = 3/t$ for {\bf \em Coord-dual}, $\alpha = 0.5$ for {\bf \em
  Coord-steep, Coord-ind}, which satisfy the condition
\eqref{eq:dim-step}, and take various values of efficiency parameter
$\beta$ from $0.5$ to $5.0$.

\smallskip
\noindent {\bf (i) Convergence to the optimal solution:} To
demonstrate our analytical findings of convergence to optimality, we
first consider simple cases which support that {\bf \em
  Coord}-algorithms find the ``accurate'' solution (\ie, the unique NE
of the game), where the exact solution can be numerically
solved. Then, we show the performance of {\bf \em Coord}-algorithms
with two cost functions, under more general topology.

\smallskip
\noindent {\bf \em Simple cases:} Let ${\bm \lambda}^\star$ and
$C^\star$ denote the optimal solution of {\bf CG-OPT} and the maximum
coordination gain of the network, respectively. We first solve the
exact optimal solution at {\tt STAR-C1}:
$\lambda_1^\star = 0.447, \ C^\star = -5.218$. Parameter updates of
node $1$ and the total coordination gain of {\bf \em Coord}-algorithms
with $\beta = 5.0$ are shown in Fig.~\ref{fig:star-s1-strategy} and
\ref{fig:star-s1-profit}, respectively. We see that all three
algorithms converge to the accurate solution after long iterations
within a range of $O(1/\beta)$ gap, yet the convergence speeds of them
do not show much difference in simple cases. Under {\tt COMP-C1}, the
exact optimal solution is attained at
$\lambda_1^\star = 0.6125, \ C^\star = -5.942$, whose convergence
results of node $1$'s activation rate and the total coordination gain
are illustrated in Fig.~\ref{fig:comp-s1-rate} and
\ref{fig:comp-s1-profit}. Note that the algorithms take shorter time
for convergence to the optimal solution in {\tt star} than in {\tt
  comp} because each node has only one edge except the hub node, \ie,
node $1$, thus pairwise interactions are less complex in {\tt star}.

\smallskip
\noindent {\bf \em Degree of coordination at optimal solution:} We
provide numerical results of two types of cost functions, under a
random topology. For both cases {\tt RAND-C1} and {\tt RAND-C2}, Fig.~\ref{fig:rand-s1-rate}, \ref{fig:rand-s2-rate} show that {\bf \em Coord-ind} with $\alpha = 0.5$, $\beta = 4.0$ converges. Note that at the
convergent status, the long-term activation rate of a node depends on
the degree of coordination, \ie, particularly in terms of (i) how many
neighbors it has, and (ii) how powerful its neighbor is. As we see in
Fig.~\ref{fig:rand-s1-rate} and \ref{fig:rand-s2-rate}, node $6$ in
Fig.~\ref{fig:rg_topology}, \ie, who has very little contribution to
the coordination gain since it has only one neighbor, achieves the
lowest long-term activation rate, while node $13$ has relatively high
long-term rate. Comparing nodes $9$ and $15$, even though both have
two neighbors, node $9$ achieves a higher long-term rate since one of
its neighbors (node $13$) is a hub so that node $9$ may implicitly
contribute to coordination gain of the network via node
$13$. Moreover, we see that the network becomes less aggressive to be
coordinated and activated if nodes have cost functions
$C_i(x) = \frac{1}{1-x}$ (\ie, under {\tt RAND-C2}), since it prevents
exclusive node activations.

\smallskip
\noindent {\bf (ii) Comparison among {\em Coord}-algorithms:}
Second, we compare the convergence of {\bf \em Coord}-algorithms under
{\tt RAND-C1}. In Fig.~\ref{fig:rand-s1-profit-comp}, we observe that
regarding the coordination gain, {\bf \em Coord-steep, Coord-ind}
converge within $10^7$ iterations, while {\bf \em Coord-dual} still
moves towards the optimal point even after $3 \times 10^7$
iterations. Note that {\bf \em Coord-dual, Coord-ind} are not designed
to follow the steepest ascent direction of {\bf CG-OPT}, thus {\bf \em
  Coord-steep} exhibits the faster convergence. 
Between {\bf \em Coord-dual} and {\bf \em Coord-ind}, we expect that
the rational and individual behavior when considering appropriate
penalty functions in \eqref{eq:penalty} brings significant
improvements in the convergence rate.

\smallskip
\noindent {\bf (iii) Trade-off between efficiency and convergence
  speed:} Finally, we present the numerical results that show the
convergence speed and efficiency (\ie, Price-of-Anarchy) of the {\bf
  \em Coord}-algorithms for various values of $\beta$. To support that
the incurring coordination gain gap due to efficiency parameter is
$1/\beta$ as stated in Theorem~\ref{thm:conv} and
Theorem~\ref{thm:NE}, we vary $\beta$ from $0.5$ to $5.0$ and plot the
coordination gain at the converged point, and measure the convergence
speed. Fig.~\ref{fig:poa} shows that, as $\beta$ grows, {\bf \em
  Coord}-algorithms require exponentially long time to converge, but
the corresponding convergent point becomes closer to the optimal
solution. From the numerical results under {\tt RAND-C1}, coordination
gain with $\beta = 4.0$ is $-21.405$ and converges after
$2.4 \times 10^7$ iterations, while that with $\beta = 0.5$ is
$-22.11$ and converges after $1.7 \times 10^5$ iterations.


%% file: conclusion.tex
\section{Conclusion and Discussion} \label{sec:conclusion}

\subsection{Summary}
In many multi-agent networked environments, a variety of gains from
coordinating actions of interacting agents are generated. In this
paper, we first formulate an optimization problem that captures the
amount of peer-to-peer coordination gain at the cost of node
activation over a given network structure, and develop three
distributed simulation-based algorithms relying only on one-hop
message passing and local observations, which we call {\bf \em
  Coord}-algorithms. It is inspired by a control of Ising model in
statistical physics, and theoretical findings of convergence to
optimality of {\bf \em Coord}-algorithms take a stochastic
approximation method that runs a Markov chain incompletely over time
with a smartly designed step-size function. We also provide new
interpretations of {\bf \em Coord-steep} and {\bf \em Coord-ind} from
a game-theoretic perspective. 

\subsection{Limitation and Future Work}
In spite of theoretical findings of convergence to optimality, our {\bf \em Coord}-algorithms may suffer from slow convergence for some dense graphs. 
Even this slow convergence issue has been observed in many prior work that use stochastic approximation theoretic update algorithms \cite{wasan:SA, kushner:SA}, there also have been
several efforts to expedite the convergence time \cite{borkar:SA,
  kesten1958sa}, which we believe, ensure practical
values of our theoretical results. Future work includes the precise analysis of the convergence rate of {\bf \em Coord}-algorithms via applying theoretical techniques, \eg, with the notion of mixing time or via weak convergence theory \cite{alex12simul, kushner77rate}.


%% file: proofs.tex
\section{Proof of Theorem~\ref{thm:conv}} \label{sec:proof1}

\subsection{Preliminary} \label{sec:preliminary} The convergence
analysis of our {\bf \em Coord}-algorithms is on the strength of
 stochastic approximation theory. As we will verify later, each of {\bf
  \em Coord}-algorithms is interpreted as a stochastic approximation
procedure with controlled continuous-time Markov process, where the
stationary distribution of the underlying Markov process from {\bf
  CDM} indeed corresponds to an Ising model. Here, we first provide
preliminary results about the convergence analysis of a general
stochastic approximation procedure with a controlled Markov process,
where an ordinary differential equation (ODE) is usefully utilized to
study the limiting behavior of the system states \cite{borkar:SA,
  borkar:noise, PY10RA}.

Consider a general discrete-time process
$\{x[t]\}_{t \in \mathbb{Z}_{\geq 0}}$ of the following form:
\begin{eqnarray} \label{eq:SA-cont} 
x[t+1] = x[t] + a[t] \cdot v(x[t], Y[t]), \quad \forall t \in \mathbb{Z}_{\geq 0},
\end{eqnarray}
where $x[t] \in \mathbb{R}^L$ is $L$-dimensional vector representing
the system state at the iteration $t$; $a[t]$ corresponds to the
step-size of the process; and $Y[t]$ is a random variable representing
the random observation (from a Markov process) during the iteration
$t$ used to update the system state. This process is often called a
{\em stochastic approximation with controlled continuous-time Markov
  process}, in \cite{borkar:SA, borkar:noise}. Here, (i)
$\{z(s)\}_{s \ge 0}$ is a stochastic process taking values in a finite
set $\mathcal{Z}$, (ii) for $s \in [t,t+1)$, $z(s)$ evolves as a
continuous-time Markov process with a control process $x[t]$, \ie,
with a controlled transition kernel $G^{x[t]}$, (iii) the observation
$Y[t]$ is a function of $\{z(s)\}_{t \le s < t+1}$, \ie,
$Y[t] = \int_t^{t+1} f(z(s)) \mathrm{d}s$, where $f(\cdot)$ is a
bounded function, and (iv) $v(x,Y)$ is a bounded, continuous,
Lipschitz in $x$ and uniformly over $Y$. We shall assume that if
$x[t]=x, \forall t$ for a fixed $x \in \mathbb{R}^L$, the controlled
Markov kernel $G^x$ is irreducible and ergodic with unique stationary
distribution $\pi^x$, and furthermore, the mapping $x \mapsto G^x$ is
continuous and $x \mapsto \pi^x$ is Lipschitz continuous. In the
following, $\xi^x(\mathrm{d}y)$ denotes the stationary distribution of
one unit iteration, \ie, $\int_0^1 f(z^x(s))\mathrm{d}s$, where
$z^x(\cdot)$ is a Markov process with $G^x$, and we also assume that
$x[t]$ remains bounded, which can be easily imposed by projecting the
process to a bounded subset of $\mathbb{R}^L$. Finally, we use a
positive monotonically decreasing step-size function $a[t]$ satisfying
\eqref{eq:dim-step}, where the example choices of such step-size
function include $a[t] = \frac{1}{t}, \frac{1}{1+t \log t}$.

Now, define a {\em virtual} time-scale
$\kappa(t) = \sum_{m=0}^{t-1} a[m]$. We take a continuous-time
piecewise linear interpolation of the system state for the time-scale
$\kappa$ in the following way: define
$\{x_{\kappa}(\tau)\}_{\tau \in \mathbb{R}_+}$ as:
$\forall t \in \mathbb{Z}_{\geq 0}$,
$\forall \tau \in [\kappa(t),\kappa(t+1))$,
\begin{eqnarray} \label{eq:SA-int}
  x_\kappa(\tau) = x[t] + (x[t+1]-x[t]) \times \frac{\tau - \kappa(t)}{\kappa(t+1) - \kappa(t)}. 
\end{eqnarray}
Intuitively, for a decreasing step-size $a[t]$, the interpolated
continuous trajectory $x_\kappa(\tau)$ is an accelerated version of
the original trajectory $x[t]$. Now, the following lemma provides the
convergence guarantee of the iterative procedure \eqref{eq:SA-cont}.

\begin{lemma} [Theorem $1$ of \cite{PY10RA}, Corollary $8$ of
  \cite{borkar:SA}(pp.74)] \label{lem:SA-cont}
  Let $T > 0$, and denote by $\tilde{x}^s(\cdot)$ the solution on
  $[s,s+T]$ of the following ODE:
\begin{eqnarray} \label{eq:SA-cont-ode}
\dot{x}(\tau) = \int_y v(x(\tau),y) \cdot \xi^{x(\tau)}(\mathrm{d}y), ~\text{ with }~ \tilde{x}^s(s) = x_\kappa(s).
\end{eqnarray}
Then, we have almost surely,
\begin{eqnarray*}
\lim_{s \rightarrow \infty} \sup_{\tau \in [s,s+T]} \| x_\kappa(\tau) - \tilde{x}^s(\tau) \| = 0.
\end{eqnarray*}
\end{lemma}
\vspace{0.05in}

Note that since the Markov process is irreducible and ergodic, and $f$
is continuous and bounded, we have,
\begin{eqnarray*}
  \int_y v(x,y) \xi^x(\mathrm{d}y) = \sum_{z \in \mathcal{Z}} v(x,f(z)) \pi^x(z), \quad \text{a.s..}
\end{eqnarray*}
Therefore, the ODE \eqref{eq:SA-cont-ode} becomes the following
simpler form, which will be used later in the proof of
Theorem~\ref{thm:conv}:
\begin{eqnarray} \label{eq:SA-cont-ode2}
\dot{x}(\tau) = \sum_{z \in \mathcal{Z}} v(x(\tau), f(z)) \pi^{x(\tau)}(z).
\end{eqnarray}

Lemma~\ref{lem:SA-cont} states that as time evolves, the dynamics of
the underlying Markov process is {\em averaged} due to the decreasing
step-size, \eg, $a[t]= \frac{1}{t}$, thus ``almost reaching the
stationary status.''  Intuitively, we expect that due to the
decreasing step-size, the speed of variations of $x[t]$ decreases and
tends to $0$ when time sufficiently grows. As consequence, the dynamic
of \eqref{eq:SA-cont} is close to that of an irreducible and ergodic
Markov process with a fixed generator (as if the system state was {\em
  frozen}), and has time to converge to its ergodic behavior. Thus, it
suffices to see how the ODE \eqref{eq:SA-cont-ode} (equivalently
\eqref{eq:SA-cont-ode2}) behaves. Moreover, when the ODE
\eqref{eq:SA-cont-ode} has a unique fixed stable equilibrium
$x^\star$, we have almost surely,
$\lim_{t \rightarrow \infty} x[t] = x^\star$.

\subsection{Proof of Theorem~\ref{thm:conv} (i): Convergence}
\label{sec:proof_conv}

We now show the convergence of {\bf \em Coord}-algorithms in
Theorem~\ref{thm:conv}. In particular, we prove that {\bf \em
  Coord}-algorithms converge to the optimal solution of the
approximated problem {\bf A-CG-OPT} in \eqref{eq:aopt}. Our main proof
strategy follows the stochastic approximation procedure whose limiting
behavior is understood by an ODE as in
Section~\ref{sec:preliminary}. For each scheme of {\bf \em
  Coord}-algorithms, the proof contains following common two steps:
first in {\bf \em Step 1}, we show that the dynamics asymptotically
approach some deterministic trajectory which is described as a
solution trajectory of an ODE system, where each scheme tracks a
slightly different deterministic trajectory. In {\bf \em Step 2}, we
then prove that the resulting deterministic trajectory converges to
the solution of {\bf A-CG-OPT}. To do this, we take the
afore-mentioned results of Lemma~\ref{lem:SA-cont} into each scheme.

\smallskip

{\bf i) \em \underline{Step 1.}} In this step, we apply preliminary
results in Section~\ref{sec:preliminary} to each scheme of {\bf \em
  Coord}-algorithms by showing that the original discrete sequence
matches with the setup defined as \eqref{eq:SA-cont} in
Section~\ref{sec:preliminary} for {\bf \em Coord-dual} or an
alternatively derived discrete sequence does for {\bf \em Coord-steep}
and {\bf \em Coord-ind}. We first verify that each scheme is a
stochastic approximation procedure with controlled Markov noise in
\eqref{eq:SA-cont}, and then provide a lemma as a direct consequence
of applying Lemma~\ref{lem:SA-cont} to our framework.

\smallskip
\noindent{\bf (a) \em Coord-dual.} To follow the analysis in
Section~\ref{sec:preliminary}, we first define a virtual time-scale
$\zeta(\cdot)$ from the step-size $a[\cdot]$ of {\bf \em Coord-dual}
as: $\zeta(t) = \sum_{m=0}^{t-1} a[m].$ We now construct
$\{{\bm \theta}(\tau)\}_{\tau \in \mathbb{R}_+}$\footnote{We omit
  $\zeta$ and use ${\bm \theta}(\tau)$ instead of
  ${\bm \theta}_\zeta(\tau)$ for notational simplicity.}, which
interpolates the discrete sequence of \eqref{eq:Q-dual} similarly to
\eqref{eq:SA-int}. We also define
$\hat{{\bm s}}(\tau) \defeq \hat{{\bm s}}[t] \cdot {\bm 1}_{\zeta(t)
  \leq \tau \leq \zeta(t+1)}$, where ${\bm 1}_{A}$ is the indicator
function for the event $A$. It then should be clear that this setup
matches with \eqref{eq:SA-cont} in Section~\ref{sec:preliminary}. The
equivalence is obtained by: $x[t] \equiv {\bm \theta}[t]$;
$Y[t] \equiv \hat{\bm s}[t]$;
$\{z(s)\}_{t \leq s <t+1} \equiv \{ {\bm \sigma}(s)\}_{t \leq s <
  t+1}$ is the process recording the configurations from
\textbf{CDM(${\bm \theta}[t]$)} during frame $t$;
$f(z(s)) \equiv {\bm \phi}({\bm \sigma}(s))$ is a coordination
configuration; $\pi^x \equiv p_{\bm \theta}$ is the stationary
distribution \eqref{eq:ccd-stationary} of the
\textbf{CDM($\bm \theta$)}; and finally
\begin{align*}
  v_i(x,y) \equiv C_i'^{-1}\bigg( \frac{-x}{\beta} \bigg) - y, \quad 
  v_{ij}(x,y) \equiv U_{ij}'^{-1}\bigg( \frac{x}{\beta} \bigg) - y.
\end{align*}
Note that under our setup of utility and cost function: strictly concave, continuously twice-differentiable utility function $U_{ij}:
  [0,1] \mapsto \mathbb{R}$ for edge $(i,j) \in E$ and strictly
  convex, continuously twice-differentiable cost function $C_i: [0,1] \mapsto
  \mathbb{R}$ for node $i$, we have followings. First, $v_{ij}(x,y):
  [\theta^{\text{min}},\theta^{\text{max}}] \times [0,1] \mapsto
  \mathbb{R}$ is Lipschitz continuous in $x$, since $U_{ij}$ is strictly convex, continuously twice-differentiable on compact set, it follows that $U_{ij}'^{-1}$ is Lipschitz continuous by the Mean Value
  Theorem. Second, $v_{ij}(x,y)$ is a linear function with respect to $y$, thus it is obvious that it is uniformly continuous in $y$. Similar arguments hold for $v_i(x,y)$. Third, Markov process generated by {\bf CDM($\bm \theta$)} is a continuous function
of $\bm \theta$, and moreover ${\bm \theta} \mapsto p_{\bm \theta}$ is
Lipschitz continuous for the bounded ${\bm \theta} \in
[\theta^{\text{min}}, \theta^{\text{max}}]$. Therefore, one can verify that the assumptions
in Section~\ref{sec:preliminary} are satisfied.

\smallskip
\noindent{\bf \em (b) Coord-steep.} 
Before to analyze the convergence of {\bf \em Coord-steep}, we provide a detail of the derivation of the rule \eqref{eq:steep-grad}. Recall that ${\bm \mu} = [\mu_{\bm \sigma}]_{{\bm \sigma} \in \set{I}(G)}$ is the probability distribution over the feasible configurations $\set{I}(G)$, and thus the definition of $\set{F}(\bm \mu)$ in \eqref{eq:opt2} can be represented in following detailed form:
\begin{align}
\set{F}(\bm \mu) &:= \sum_{(i,j) \in E} U_{ij}\big( \mathbb{E}_{\bm \mu}[\sigma_i \sigma_j] \big) - \sum_{i \in V} C_i( \mathbb{E}_{\bm \mu}[\sigma_i] ) \cr 
&= \sum_{(i,j) \in E} U_{ij}\big( \sum_{{\bm \sigma} \in \set{I}(G)} \mu_{\bm \sigma} \cdot \sigma_i \sigma_j \big) - \sum_{i \in V} C_i \big( \sum_{{\bm \sigma} \in \set{I}(G)} \mu_{\bm \sigma} \cdot \sigma_i \big).
\end{align}
Now, using the chain rule, a partial derivative of $\set{F}(\bm \mu)$ with respect to the variable $\mu_{\bm \sigma}$ is derived as follows:
\begin{align}
\frac{\partial \set{F}(\bm \mu)}{\partial \mu_{\bm \sigma}} &= \sum_{(i,j) \in E} U_{ij}'\big( \mathbb{E}_{\bm \mu}[\sigma_i \sigma_j] \big) \cdot \frac{\partial \mathbb{E}_{\bm \mu}[\sigma_i \sigma_j] }{\partial \mu_{\bm \sigma}} - \sum_{i \in V} C_i'\big( \mathbb{E}_{\bm \mu}[\sigma_i] \big) \cdot \frac{\partial \mathbb{E}_{\bm \mu}[\sigma_i]}{\partial \mu_{\bm \sigma}} \cr 
&= \sum_{(i,j) \in E} U_{ij}'\big( \mathbb{E}_{\bm \mu}[\sigma_i \sigma_j] \big) \cdot \sigma_i \sigma_j - \sum_{i \in V} C_i'\big( \mathbb{E}_{\bm \mu}[\sigma_i] \big) \cdot \sigma_i.
\end{align}

Now, the first step is to approximate
{\bf \em Coord-steep} for large $t$ by the dynamic of a
continuous-time ODE system, by taking a continuous-time
interpolation. While a stochastic approximation idea in {\bf \em
  Coord-dual} comes from the diminishing step-size $a[\cdot]$,
adopting the cumulative rate $\bar{\bm s}[t]$ in {\bf \em Coord-steep}
plays the similar role (see the relation
\eqref{eq:serv_rate}). 
To understand the role of $\bar{\bm s}[t]$, we introduce an {\em alternative} discrete-time sequence
$\{{\bm \rho}[t]\}_{t \in \mathbb{Z}_{\geq 0}}$ derived from
$\{{\bm \theta}[t]\}_{t \in \mathbb{Z}_{\geq 0}}$ of {\bf \em
  Coord-steep} in \eqref{eq:Q-steep} defined as:
\begin{eqnarray*} 
  {\bm \rho}[t] &=& \frac{1}{\alpha} \cdot {\bm \theta}[t] + \Big( 1- \frac{1}{\alpha} \Big) \cdot {\bm \theta}[t-1], 
\end{eqnarray*}
and thus we have the following property by applying recursion:
\begin{eqnarray} \label{eq:r_dif_jd}
  {\bm \theta}[t] &=& \alpha \cdot {\bm \rho}[t] + (1-\alpha) \cdot {\bm \theta}[t-1] \cr
  &=& \alpha {\bm \rho}[t] + (1-\alpha) \Big( \alpha {\bm \rho}[t-1] + (1-\alpha) {\bm \theta}[t-2] \Big) \cr
      &=& \cdots \ = \sum_{m=0}^{t-1} \alpha (1-\alpha)^m {\bm \rho}[t-m].   
\end{eqnarray}
Then, from \eqref{eq:Q-steep} and \eqref{eq:r_dif_jd}, {\bf \em
  Coord-steep} can be understood as the update rule
$\{{\bm \rho}[t]\}_{t \in \mathbb{Z}_{\geq 0}}$ of following form:
\begin{eqnarray} \label{eq:Q-aJD}
  \rho_i[t+1] = -\beta C_i'(\bar{s}_i[t]), \quad 
  \rho_{ij}[t+1] = \beta U_{ij}'(\bar{s}_{ij}[t]),
\end{eqnarray}
and thus we have 
\begin{align*}
\bar{s}_i[t-1] = C_i'^{-1}\bigg(\frac{-\rho_i[t]}{\beta}\bigg), \quad \bar{s}_{ij}[t-1] = U_{ij}'^{-1} \bigg(\frac{\rho_{ij}[t]}{\beta}\bigg).
\end{align*}
Now, when $t$ grows large, the update rule \eqref{eq:Q-aJD} becomes
approximately as follows under the assumption {\bf (A1)},
\begin{align} \label{eq:SAmarkov_JD}
  \rho_i[t+1] & = - \beta C_i'(\bar{s}_i[t]) \cr
& \stackrel{(a)}{=} - \beta \Big( C_i'(\bar{s}_i[t-1]) + \frac{1}{t}(\hat{s}_i[t] - \bar{s}_i[t-1]) C_i''( \bar{s}_i[t-1]) \Big) \cr
&= -\beta C_i'(\bar{s}_i[t-1]) + \frac{1}{t} g_i(\rho_i[t]) (\bar{s}_i[t-1] - \hat{s}_i[t]) \cr
  & \stackrel{(b)}{=} \rho_i[t] + \frac{1}{t} g_i(\rho_i[t]) \Big( C_i'^{-1} \Big( \frac{-\rho_i[t]}{\beta} \Big) - \hat{s}_i[t] \Big),
\end{align}
and similarly,
\begin{align*}
& \rho_{ij}[t+1] = \rho_{ij}[t] + \frac{1}{t} g_{ij}(\rho_{ij}[t]) \Big( U_{ij}'^{-1} \Big( \frac{\rho_{ij}[t]}{\beta} \Big) - \hat{s}_{ij}[t] \Big),
\end{align*}
where $g_i(\cdot)$, $g_{ij}(\cdot)$ are defined in \eqref{eq:g-fun},
and both are positive for convex, increasing function $C_i(\cdot)$ and
concave, increasing function $U_{ij}(\cdot)$. The equality $(a)$ holds
from a first-order Taylor's expansion, $(b)$ comes from the {\bf
  \em Coord-steep} rule.

To take afore-mentioned results in Section~\ref{sec:preliminary} into
this framework, we define another virtual time-scale $\kappa(t)$ as
$\kappa(t) \defeq \sum_{m=1}^{t-1} \frac{1}{m}$ with $\kappa(0) = 0$, since the discrete sequence of {\bf \em
  Coord-steep} is interpreted to have a step-size $\frac{1}{t}$ at
iteration $t$, see \eqref{eq:SAmarkov_JD}. We construct an
interpolated trajectory $\{{\bm \rho}(\tau)\}_{\tau \in \mathbb{R}_+}$
from the discrete sequence
$\{{\bm \rho}[t]\}_{t \in \mathbb{Z}_{\geq 0}}$ in \eqref{eq:Q-aJD}
with time-scale $\kappa$. Then, it should be clear that the
alternative process matches with the setup \eqref{eq:SA-cont} in
Section~\ref{sec:preliminary}. The equivalence is obtained by
$x[t] \equiv {\bm \rho}[t]$; $Y[t] \equiv \hat{\bm s}[t]$;
$a[t] = \frac{1}{t}$;
$\{z(s)\}_{t \leq s <t+1} \equiv \{ {\bm \sigma}(s)\}_{t \leq s <
  t+1}$; $f(z(s)) \equiv {\bm \phi}({\bm \sigma}(s))$;
$\pi^x \equiv p_{\bm \theta}$ in \eqref{eq:ccd-stationary}; and
finally
\begin{align*}
  v_i(x,y) &\equiv g_i(x) \Big( C_i'^{-1}\Big( \frac{-x}{\beta} \Big) - y \Big), \quad i \in V, \cr
  v_{ij}(x,y) &\equiv g_{ij}(x) \Big( U_{ij}'^{-1}\Big( \frac{x}{\beta} \Big) - y \Big), \quad (i,j) \in E.
\end{align*}

\smallskip
\noindent{\bf (c) \em Coord-ind.} Here, we also introduce an
alternative discrete-time sequence
$\{{\bm \eta}[t]\}_{t \in \mathbb{Z}_{\geq 0}}$ derived from
$\{{\bm \theta}[t]\}_{t \in \mathbb{Z}_{\geq 0}}$ of {\bf \em
  Coord-ind} in \eqref{eq:Q-ind} as follows:
\begin{align*}
  {\bm \eta}[t] = \frac{1}{{\bm \gamma}[t-1]} \cdot {\bm \theta}[t] + \Big( 1 - \frac{1}{{\bm \gamma}[t-1]} \Big) \cdot {\bm \theta}[t-1],
\end{align*}
where ${\bm \gamma}[t]$ is given by 
\begin{align} \label{eq:ind-gamma}
  \gamma_i[t] = \frac{\alpha}{\beta}\frac{\partial s_{i}({\bm \theta}[t])}{\partial \theta_{i}}, \ \
\gamma_{ij}[t] = \frac{\alpha}{\beta}\frac{\partial s_{ij}({\bm \theta}[t])}{\partial \theta_{ij}}.
\end{align}
Then, we have the following property by applying recursion:
\begin{align} \label{eq:r_dif_gd}
  {\bm \theta}[t] &= {\bm \gamma}[t-1] \cdot {\bm \eta}[t] + (1 - {\bm \gamma}[t-1]) \cdot {\bm \theta}[t-1] \cr
  &= {\bm \gamma}[t-1] {\bm \eta}[t] \cr
   & ~\mbox{} \quad \ + \sum_{m=1}^{t-1} \prod_{l=1}^m \Big(1-{\bm \gamma}[t-l]\Big) {\bm \gamma}[t-m-1] {\bm \eta}[t-m].
\end{align}
Now, from \eqref{eq:Q-ind} and \eqref{eq:r_dif_gd}, {\bf \em
  Coord-ind} can be understood as the update rule
$\{{\bm \eta}[t]\}_{t \in \mathbb{Z}_{\geq 0}}$ of representation in
\eqref{eq:Q-aGD}.
Then, two sequences $\{{\bm \rho}[t]\}_{t \in \mathbb{Z}_{\geq 0}}$ in
\eqref{eq:Q-aJD} and $\{{\bm \eta}[t]\}_{t \in \mathbb{Z}_{\geq 0}}$
in \eqref{eq:Q-aGD} are evolved in the same way. Therefore, an
interpolated trajectory $\{{\bm \eta}(\tau)\}_{\tau \in \mathbb{R}_+}$
from $\{{\bm \eta}[t]\}_{t \in \mathbb{Z}_{\geq 0}}$ with time-scale
$\kappa$ is equivalent to
$\{{\bm \rho}(\tau)\}_{\tau \in \mathbb{R}_+}$.

\smallskip

From the equivalence above, following is a direct consequence from
Lemma~\ref{lem:SA-cont} in Section~\ref{sec:preliminary}, which states
that the interpolated trajectory of each scheme asymptotically tracks
the solution trajectory of the corresponding ODE system.

\begin{lemma} \label{lem:averaging} Let $T >0$, and fix $w > 0$. 
  \smallskip
  \noindent {\bf (i) {\em Coord-dual}.} Denote by $\tilde{\bm \theta}^w(\cdot)$
  the solution on $[w,w+T]$ of the following ODE: $\forall i \in V$
  and $\forall (i,j) \in E$,
  \begin{align} \label{eq:ode_max} 
  \dot{\theta_i}(\tau) &= C_i'^{-1}
    \bigg( \frac{-\theta_i(\tau)}{\beta}\bigg) - 
    \mathbb{E}_{{{\bm \theta}(\tau)}}[\sigma_i], \cr
    \dot{\theta_{ij}}(\tau) &= U_{ij}'^{-1} \bigg(
    \frac{\theta_{ij}(\tau)}{\beta} \bigg) - 
    \mathbb{E}_{{{\bm \theta}(\tau)}}[\sigma_i \sigma_j],
  \end{align}
with $\tilde{{\bm \theta}}^w(w) = {\bm \theta}(w)$. Then, we have almost surely, 
  \begin{eqnarray*}
  \lim_{w \rightarrow \infty} \sup_{\tau \in [w,w+T]} \| {\bm \theta}(\tau) - \tilde{\bm \theta}^w(\tau) \| = 0.
  \end{eqnarray*}

\smallskip
\noindent {\bf (ii) {\em Coord-steep}} and {\bf {\em Coord-ind}.}
Denote by $\tilde{\bm{\rho}}^w(\cdot)$ the solution on $[w,w+T]$ of
the following ODE: $\forall i \in V$ and $\forall (i,j) \in E$,
  \begin{align} \label{eq:ode_JD} 
    \dot{\rho}_i(\tau) &=
       g_i\Big(\rho_i(\tau)\Big) \bigg[C_i'^{-1}\bigg(
      \frac{-\rho_i(\tau)}{\beta} \bigg) - 
      \mathbb{E}_{{{\bm \theta}(\tau)}}[\sigma_i] \bigg],
      \cr 
  \dot{\rho}_{ij}(\tau) &= g_{ij}\Big(\rho_i(\tau)\Big)
      \bigg[U_{ij}'^{-1}\bigg( \frac{\rho_{ij}(\tau)}{\beta}
      \bigg) - 
      \mathbb{E}_{{{\bm \theta}(\tau)}}[\sigma_i \sigma_j] \bigg], 
  \end{align}
  with $\tilde{\bm \rho}^w(w) = {\bm \rho}(w) = {\bm \eta}(w)$. Then,
  we have almost surely,
  \begin{eqnarray*}
    \lim_{w \rightarrow \infty} \sup_{\tau \in [w,w+T]} \| {\bm \rho}(\tau) - \tilde{\bm \rho}^w(\tau) \| = 0, 
  \end{eqnarray*}
  and equivalently,
  \begin{eqnarray*}
    \lim_{w \rightarrow \infty} \sup_{\tau \in [w,w+T]} \| {\bm \eta}(\tau) - \tilde{\bm \rho}^w(\tau) \| = 0.
  \end{eqnarray*}  
\end{lemma}

\smallskip
{\bf ii) \em \underline{Step 2.}} Now, we prove that each ODE system
in Lemma~\ref{lem:averaging} has a unique fixed point, and thus the
resulting deterministic solution trajectory converges to the point. We
then show that the point attains at the optimal solution of {\bf
  A-CG-OPT}, \ie, {\bf \em Coord}-algorithms converge to
${\bm \theta}^\circ$.

\smallskip
\noindent{\bf (a) {\em Coord-dual}.} We show that the ODE system
\eqref{eq:ode_max} has the solution of {\bf A-CG-OPT}, denoted by
${\bm \theta}^\circ$, as a unique fixed point. In particular,
\eqref{eq:ode_max} may be interpreted as a sub-gradient dynamics
solving the {\em dual} of the convex problem {\bf A-CG-OPT}. To that
end, we first consider the Lagrangian $\set{L}$ of {\bf A-CG-OPT} with
dual variables ${\bm k} = ([k_i]_{i \in V}, [k_{ij}]_{(i,j) \in E})$:
\begin{align} \label{eq:lagrange}
   \set{L}({\bm \mu},{\bm \lambda}; {\bm k})
& = \sum_{(i,j) \in E}
  U_{ij}(\lambda_{ij}) - \sum_{i \in V} C_i(\lambda_i) + \frac{1}{\beta}H(\bm \mu) \cr
& ~\mbox{} \quad + \sum_{i \in V} k_i (\mathbb{E}_{\bm \mu}[\sigma_i] - \lambda_i) 
 + \sum_{(i,j) \in E} k_{ij} (\mathbb{E}_{\bm \mu}[\sigma_i\sigma_j] - \lambda_{ij}). 
\end{align}
The primal solution of {\bf A-CG-OPT} is the minimum point of the dual
function, which is given by
$\set{D}({\bm k})= \sup_{{\bm \mu},{\bm \lambda}} \set{L}({\bm
  \mu},{\bm \lambda};{\bm k}).$
Finally, the dual problem is formulated as:
\begin{eqnarray} \label{eq:dual}
\min_{{\bm k} \in \mathbb{R}^{|V|+|E|}} ~ \set{D}(\bm k).
\end{eqnarray}
Note that the primal problem in \eqref{eq:aopt} is a concave
maximization and the dual problem in \eqref{eq:dual} is a convex
minimization from the concavity of entropy and $U_{ij}(\cdot)$,
convexity of $C_i(\cdot)$ under our setup. Therefore, following the
results in standard primal-dual optimization theory, there is no
duality gap and both have the same, unique solution, and moreover its
sub-gradient algorithm will converge to the solution.

Given a feasible dual variable ${\bm k}$, let
$(\bm{\mu}^\dagger(\bm k), {\bm \lambda}^\dagger(\bm k))$ be the
corresponding feasible primal solution that maximizes the Lagrangian
$\set{L}$. Given the structure of $\set{L}$ in \eqref{eq:lagrange},
from Karush-Kuhn-Tucker (KKT) conditions of {\bf A-CG-OPT}, it follows
that ${\bm \mu}^\dagger(\bm k), {\bm \lambda}^\dagger({\bm k})$ should
be such that:
\begin{eqnarray} \label{eq:kkt-lambda}
&&{\bm \mu}^\dagger_{\bm \sigma}(\bm k) \propto \exp \Big(\sum_{i \in V} \beta k_i \sigma_i + \sum_{(i,j) \in E} \beta k_{ij} \sigma_i\sigma_j \Big), ~\quad \forall {\bm \sigma} \in \set{I}(G), \cr
&&\lambda^\dagger_i(\bm k) = ~\text{arg} \max_{y \in [0,1]} \Big[ -
                            C_i(y) - k_i y \Big], ~\quad \forall i \in V, \cr
&&\lambda^\dagger_{ij}(\bm k) = ~\text{arg} \max_{y \in [0,1]}
                              \Big[ U_{ij}(y) - k_{ij} y \Big], ~\quad \forall (i,j) \in E.
\end{eqnarray}

Now, we can conclude that
${\bm \mu}^\dagger(\bm \theta) = p_{\bm \theta}$ with
${\bm \theta} = \beta {\bm k}$, from \eqref{eq:ccd-stationary} and
\eqref{eq:kkt-lambda}. Then, the dual function is represented with
respect to ${\bm \theta}$ as
$\set{D}(\bm \theta) = \set{L}({\bm \mu}^\dagger(\bm \theta), {\bm
  \lambda}^\dagger(\bm \theta); {\bm \theta}),$ 
and the slack in each constraint is given by:
\begin{align} \label{eq:kkt-dual}
\mathbb{E}_{{\bm \mu}^\dagger(\bm \theta)}[\sigma_i] - \lambda_i^\dagger(\bm \theta), \quad ~\text{and}~ \quad
\mathbb{E}_{{\bm \mu}^\dagger(\bm \theta)}[\sigma_i \sigma_j] - \lambda_{ij}^\dagger(\bm \theta).
\end{align}
Accounting for \eqref{eq:kkt-lambda} and \eqref{eq:kkt-dual}, the
sub-gradient algorithm solving the dual problem \eqref{eq:dual} with
parameter ${\bm \theta}$, \ie, using $\grad \set{D}({\bm \theta})$, is
given by following ODEs: $\forall i \in V, ~ \forall (i,j) \in E$,
\begin{eqnarray} \label{eq:subgrad}
    \dot{\theta_i} = C_i'^{-1} \bigg(\frac{-\theta_i}{\beta} \bigg) - \mathbb{E}_{{\bm \theta}}[\sigma_i], ~\mbox{} \
   \dot{\theta_{ij}} = U_{ij}'^{-1} \bigg(\frac{\theta_{ij}}{\beta} \bigg) - \mathbb{E}_{{\bm \theta}}[\sigma_i \sigma_j].
\end{eqnarray}
which is obviously equivalent to \eqref{eq:ode_max}, provided that
${\bm \theta}(\tau)$ remains between
$[\theta^{\text{min}}, \theta^{\text{max}}]$ component-wisely. Note
that the dual solution, denoted by ${\bm \theta}^\circ$, actually
belongs to the interval $[\theta^{\text{min}},\theta^{\text{max}}]$
component-wisely, as a fixed point of \eqref{eq:subgrad}, under {\bf
  (A1)}. Therefore, the sub-gradient converges to the dual solution
${\bm \theta}^\circ$, where the corresponding primal solution is
$({\bm \mu}^\circ, {\bm \lambda}^\circ)$\footnote{It is obvious that
  we mean
  ${\bm \mu}^\circ = {\bm \mu}^\dagger({\bm \theta}^\circ) = p_{{\bm
      \theta}^\circ}$ and
  ${\bm \lambda}^\circ = {\bm \lambda}^\dagger({\bm \theta}^\circ)$.},
and hence the solution trajectory of the ODE system \eqref{eq:ode_max}
also does. Finally, we can conclude that under {\bf \em Coord-dual},
we have almost surely,
$\lim_{t \rightarrow \infty} {\bm \theta}[t] = {\bm \theta}^\circ.$

\smallskip
\noindent {\bf (b) {\em Coord-steep}.} In case of {\bf \em
  Coord-steep}, we need an additional step that proves the equivalence
of the convergence of alternative process
$\{{\bm \rho}[t]\}_{t \in \mathbb{Z}_{\geq 0}}$ and that of
$\{ {\bm \theta}[t]\}_{t \in \mathbb{Z}_{\geq 0}}$. Since parameter
lies in compact region, from \eqref{eq:r_dif_jd}, there exist
constants $L$ and $M$ such that $\forall t \in \mathbb{Z}_{\geq 0}$,
\begin{align} \label{eq:steep-const}
  \left|\rho_i [t] \right| < L, \ \ \left|g_i(\rho_i[t]) \Big(
    C_i'^{-1} \Big( \frac{-\rho_i[t]}{\beta} \Big) - \hat{s}_i[t]
    \Big) \right| < M.
\end{align}
For $\epsilon > 0$, let
$T(\epsilon) \defeq \frac{4 \log(\frac{\epsilon M}{4L})}{\epsilon
  \log(1-\alpha)}$. Then, for all $t \geq T(\epsilon)$,
$ \left| \rho_i [t] - \theta_i[t] \right| \le \frac{5}{4} \epsilon M,$
because\footnote{Here, we use
  just $\epsilon t/4$ instead of $\lceil \epsilon t/4 \rceil$ for
  notionally simplicity.}
\begin{align*}
  \left| \rho_i [t] -\theta_i[t]\right| & \stackrel{(a)}{=} 
  \left| \rho_i[t] - \sum_{m=0}^{t-1} \rho_i
      [t-m] \alpha(1-\alpha)^m\right| \cr &  \le 
  \sum_{m=0}^{t-1}\left| \rho_i[t] - \rho_i
      [t-m]\right| \alpha(1-\alpha)^m + \left|\rho_i[t]\right| (1-\alpha)^t \cr
  &  \stackrel{(b)}{\le}  \frac{\epsilon t /4 }{ t- \epsilon t /4}M + 2L \sum_{m=
    \epsilon t /4} ^{t-1} \alpha (1-\alpha)^m + L (1-\alpha)^{\epsilon t/4} \cr
  & \stackrel{(c)}{\le}  \frac{\epsilon}{2} M + 2L (1-\alpha)^{\epsilon t /4} + L (1-\alpha)^{\epsilon t/4} ~\stackrel{(d)}{\le}~ \frac{5}{4} \epsilon M,
\end{align*}
where $(a)$ comes from \eqref{eq:r_dif_jd}, $(b)$ and $(c)$ come from
the following using triangle inequality and \eqref{eq:SAmarkov_JD},
\eqref{eq:steep-const}:
\begin{align*}
  \sum_{m=0}^{\epsilon t/4-1} & \left| \rho_i [t] - 
    \rho_i[t-m]\right| \alpha(1-\alpha)^m \cr &~\le~
  \sum_{m=1}^{\epsilon t/4-1}\sum_{k=1}^{m}\left| \rho_i[t-k+1] - \rho_i [t-k]\right|
  \alpha(1-\alpha)^m \cr & ~~\le~ \sum_{m=1}^{\epsilon
    t/4-1}\frac{m\cdot M}{t-m} \alpha(1-\alpha)^m ~\le~
  \frac{\epsilon t /4 }{ t- \epsilon t /4}M,
\end{align*}
and finally $(d)$ holds for $t \geq T(\epsilon)$ and
$\epsilon \leq 2$. Therefore, we have the following relation between
${\bm \rho}[t]$ and ${\bm \theta}[t]$:
\begin{eqnarray} \label{eq:nu-theta}
\lim_{t \rightarrow \infty} {\bm \rho}[t] - {\bm \theta}[t] = 0.
\end{eqnarray}

Now, we observe that the ODE system \eqref{eq:ode_JD} is equivalent to
\eqref{eq:ode_max}, because we have positive
$g_i(\cdot), g_{ij}(\cdot)$. Therefore, the ODE system
\eqref{eq:ode_JD} converges to a unique fixed point, say
${\bm \rho}^{\circ}$, such that
${\bm \theta}^\circ = {\bm \rho}^{\circ}$. Finally, from
\eqref{eq:nu-theta}, we can conclude that under {\bf \em Coord-steep},
we have almost surely,
$\lim_{t \rightarrow \infty} {\bm \theta}[t] = {\bm \theta}^\circ$.

\smallskip
\noindent{\bf (c) {\em Coord-ind}.} In case of {\bf \em Coord-ind},
similarly to the earlier scheme {\bf \em Coord-steep}, we prove the
equivalence of the convergence of
$\{{\bm \eta}[t]\}_{t \in \mathbb{Z}_{\geq 0}}$ and that of
$\{{\bm \theta}[t]\}_{t \in \mathbb{Z}_{\geq 0}}$. From the simple
algebra of \eqref{eq:marginal}, we have component-wisely:
\begin{eqnarray} \label{eq:s_der}
  \grad {\bm s}(\bm \theta) = {\bm s}(\bm \theta) \Big( 1 - {\bm s}({\bm \theta}) \Big).
\end{eqnarray}
We first denote by $\gamma^{\text{min}}$ and $\gamma^{\text{max}}$ the
minimum and maximum value of the sequence
$\{{\bm \gamma}[t]\}_{t \in \mathbb{Z}_{\geq 0}}$ in
\eqref{eq:ind-gamma}, respectively. Note that for sufficiently large
$\beta$, $\gamma^{\text{min}}, \gamma^{\text{max}}$ is less than $1$,
since $\grad {\bm s}({\bm \theta}) \in (0,1/4)$ from
\eqref{eq:s_der}. Now, since parameter ${\bm \theta}$ lies in compact
region, from \eqref{eq:r_dif_gd}, there also exist constants $L$ and
$M$ such that $\forall t \in \mathbb{Z}_{\geq 0}$,
\begin{align*}
  \left|\eta_i [t] \right| < L, \quad ~\text{and}~ \quad \left| g_i(\eta_i[t]) \Big(
    C_i'^{-1} \Big( \frac{-\eta_i[t]}{\beta} \Big) - \hat{s}_i[t] \Big) \right| < M.
\end{align*}
For $\epsilon >0$, let $S(\epsilon) \defeq \frac{4 \log ( \min ( \frac{\epsilon M}{4L} ,
  (\frac{\epsilon M}{4L} -
  \frac{\gamma^{\text{min}}}{\gamma^{\text{max}}} ) \cdot
  (\frac{\gamma^{\text{max}}}{\gamma^{\text{min}}})^2 ) ) }{\epsilon
  \log(1-\gamma^{\text{min}})}$. Then, for all $t \geq S(\epsilon)$
and $\epsilon \le 2$, similar argument in {\bf \em Coord-steep} can be
done to show
$\left| \eta_i[t] - \theta_i[t] \right| \leq
\frac{5}{4}\frac{\gamma^{\text{max}}}{\gamma^{\text{min}}}\epsilon M$.
Now, we have the following relation between ${\bm \eta}[t]$ and
${\bm \theta}[t]$ that:
\begin{eqnarray} \label{eq:upsilon-theta}
\lim_{t \rightarrow \infty} {\bm \eta}[t] - {\bm \theta}[t] = 0.
\end{eqnarray}
From the equivalence of
$\{{\bm \rho}[t]\}_{t \in \mathbb{Z}_{\geq 0}}$ and
$\{{\bm \eta}[t]\}_{t \in \mathbb{Z}_{\geq 0}}$, combined with
Lemma~\ref{lem:averaging} and the argument of the ODE system
\eqref{eq:ode_JD} in {\bf \em Coord-steep}, we can conclude that under
{\bf \em Coord-ind}, we have almost surely,
$\lim_{t \rightarrow \infty} {\bm \theta}[t] = {\bm \theta}^\circ$.

\smallskip
Consequently, from {\bf \em Step 1} and {\bf \em Step 2}, we complete
the proof that under all {\bf \em Coord}-algorithms, ${\bm \theta}[t]$
converges to ${\bm \theta}^\circ$, where the optimal solution of {\bf
  A-CG-OPT} is attained. $\qed$

\subsection{Proof of Theorem~\ref{thm:conv} (ii):  Optimality}
\label{sec:proof_opt}

We now show the asymptotic optimality of {\bf \em Coord}-algorithms by
verifying the relation between the solution of {\bf A-CG-OPT} and that
of {\bf CG-OPT}. In Section~\ref{sec:proof_conv}, we have shown that
the result of {\bf \em Coord}-algorithms converges to
${\bm \theta}^\circ$ such that the corresponding
$(p_{{\bm \theta}^\circ}, {\bm s}({\bm \theta}^\circ))$ is the
solution of {\bf A-CG-OPT} in \eqref{eq:aopt}. To establish a goodness
of the result of {\bf \em Coord}-algorithms, note that the optimal
solution of {\bf CG-OPT} in \eqref{eq:opt}, denoted by
${\bm \lambda}^\star$, along with an appropriate distribution
${\bm \mu}^\star \in \set{M}$, is one feasible solution of the problem
{\bf A-CG-OPT}. Therefore, it follows that
\begin{align*}
  \sum_{(i,j) \in E}  U_{ij}(\lambda_{ij}^\star) &- \sum_{i \in V} C_i(\lambda_i^\star)  \\
  &\stackrel{(a)}{\leq}  \sum_{(i,j) \in E} U_{ij}(\lambda_{ij}^\star) - \sum_{i \in V} C_i(\lambda_i^\star) + \frac{1}{\beta}H({\bm \mu}^\star) \\
  &\stackrel{(b)}{\leq} \sum_{(i,j) \in E} U_{ij}({\lambda}^\circ_{ij}) - \sum_{i \in V} C_i({\lambda}^\circ_i) + \frac{1}{\beta}H({\bm \mu}^\circ) \\
  &\stackrel{(c)}{\leq} \sum_{(i,j) \in E} U_{ij}({\lambda}^\circ_{ij}) - \sum_{i \in V} C_i({\lambda}^\circ_i) + \frac{\log |\set{I}(G)|}{\beta}.
\end{align*}
In the above, the first inequality $(a)$ comes from the fact that the
entropy is non-negative, $(b)$ holds since
$({\bm \mu}^\circ, {\bm \lambda}^\circ)$ is the optimal solution of {\bf
  A-CG-OPT}, and finally we have used the fact that the maximum value
of a discrete-valued random variable's entropy is at most the
logarithm of the cardinality of the support set $|\set{I}(G)|$, for
the last inequality $(c)$. $\qed$


%% file: appendix.tex
\section{Proof of Theorem~\ref{thm:NE}} 
\renewcommand{\thesubsection}{\Alph{subsection}}
\label{sec:proof2}

\begin{proof}
  \noindent{\bf \em (i) \underline{Uniqueness.}} We prove the
  existence and the uniqueness of non-trivial NE in our game
  $\text{CoordGain}(\beta)$ using a potential game approach. Consider
  the following function $P({\bm \theta})$ on the space
  $\set{C}^+ = \{ {\bm \theta}: {\bm s}({\bm \theta})>0\}$, \ie, the
  set of strategies producing ``non-trivial'' rates, defined by:
  \begin{eqnarray} \label{eq:potential} P({\bm \theta}) \defeq -
    \sup_{{\bm \mu} \in \set{M},{\bm \lambda}\in [0,1]^{|V|+|E|}} \set{L}({\bm \mu},{\bm \lambda};\frac{\bm
      \theta}{\beta}),
  \end{eqnarray}
  where $\set{L}(\cdot)$ is defined in \eqref{eq:lagrange} to describe
  a dual function of {\bf A-CG-OPT}. It is easy to check that
  $P({\bm \theta})$ is strictly concave in ${\bm \theta}$ since
  $P(\cdot)$ is the supremum of $\set{L}(\cdot)$, which is a family of
  affine functions in $\bm \theta$. We now show that
  $\text{CoordGain}(\beta)$ for any constant $\beta > 0$ is an {\em
    ordinal potential game} \cite{mon96potential} with the potential
  function $P({\bm \theta})$, \ie,
  $\text{sgn}\frac{\partial \Psi_n({\bm \theta})}{\partial \theta_n} =
  \text{sgn}\frac{\partial P(\bm{\theta})}{\partial \theta_n},$ for
  all players $n \in N.$ To verify this, we first have derivative of
  each player's payoff function: for each node player $i \in V$,
  \begin{eqnarray*}
    \frac{\partial \Psi_i({\bm \theta})}{\partial \theta_i} 
    &=& - \frac{\partial}{\partial \theta_i} \bigg( C_i(s_i({\bm \theta})) + \frac{1}{\beta}\int_{-\infty}^{\theta_i} x s_i'(x, \theta_{-i}) \mathrm{d}x \bigg) \cr
    &=& - \frac{\partial s_i({\bm \theta})}{\partial \theta_i}\bigg( C_i'(s_i({\bm \theta})) + \frac{\theta_i}{\beta} \bigg)\cr
    &=& - s_i({\bm \theta})\Big( 1 - s_i({\bm \theta}) \Big)\Big( 
        C_i'(s_i({\bm \theta})) + \frac{\theta_i}{\beta}\Big), 
  \end{eqnarray*}
  where the last equality comes from \eqref{eq:s_der}. Similarly, we
  also have for each edge player $(i,j) \in E$,
  \begin{eqnarray*}
    \frac{\partial \Psi_{ij}({\bm \theta})}{\partial \theta_{ij}} = s_{ij}({\bm \theta})\Big( 1 - s_{ij}({\bm \theta}) \Big)\Big( 
    U_{ij}'(s_{ij}({\bm \theta})) - \frac{\theta_{ij}}{\beta}\Big).
  \end{eqnarray*}
  Second, we have derivative of a potential function as:
  \begin{align*}
\frac{\partial P(\bm{\theta})}{\partial \theta_i} = C_i'^{-1}\Big( \frac{- \theta_i}{\beta}\Big) - s_i({\bm \theta}), \quad ~\text{and}~ \quad
\frac{\partial P(\bm{\theta})}{\partial \theta_{ij}} = U_{ij}'^{-1}\Big( \frac{\theta_{ij}}{\beta}\Big) - s_{ij}({\bm \theta}).
  \end{align*}
  Therefore, on the space $\set{C}^+$,
  $\text{sgn}\frac{\partial \Psi_n({\bm \theta})}{\partial \theta_n} =
  \text{sgn}\frac{\partial P(\bm{\theta})}{\partial \theta_n}$ for
  each player $n \in N$. From the standard results in potential game
  theory and strict concavity of $P(\cdot)$, the solution that
  maximizes $P(\cdot)$ is a NE ${\bm \theta}^{\text{NE}}$, where each
  player's strategy is a best response to the others' strategies, \ie,
  $\grad \Psi_n({\bm \theta}) = 0, \forall n \in N$, and moreover it
  is unique and non-trivial.

  \smallskip
  \noindent{\bf \em (ii) \underline{Price-of-Anarchy.}} The proof of
  Price-of-Anarchy follows the same argument in
  Section~\ref{sec:proof_opt}, since we observe that the unique
  non-trivial NE in our game ${\bm \theta}^{\text{NE}}$ coincides with
  the optimal solution of {\bf A-CG-OPT}, \ie,
  ${\bm \theta}^{\text{NE}} = {\bm \theta}^\circ$. From the analysis
  in Section~\ref{sec:proof_opt}, we can easily verify that
  \begin{align*}
    \text{PoA} &= \frac{\sum_{(i,j) \in E} U_{ij}(s_{ij}({\bm \theta}^\star)) - \sum_{i \in V} C_i(s_i({\bm \theta}^\star))}{\sum_{(i,j) \in E} U_{ij}(s_{ij}({\bm \theta}^{\text{NE}})) - \sum_{i \in V} C_i(s_i({\bm \theta}^{\text{NE}}))} \cr
    &= 1 + O\bigg(\frac{\log |\set{I}(G)|}{\beta}\bigg),
  \end{align*}
  and thus we have $\lim_{\beta \rightarrow \infty} \text{PoA} = 1$.
\end{proof}
